\documentclass[a4paper,10pt]{article}

\usepackage{amssymb}
\usepackage{amsfonts}
\usepackage{amsmath}
\usepackage{subfigure}
\usepackage{algorithm}
\usepackage[noend]{algorithmic}
\usepackage{amsmath,amssymb,cite}
\usepackage{epsfig}
\usepackage{pst-node,pstricks}

\usepackage{diagbox}
\usepackage{array}
\usepackage{tabu}
\usepackage{multirow}
\usepackage{pict2e}
\usepackage{slashbox,booktabs,amsmath}
\usepackage{diagbox}
\usepackage{enumitem}
\usepackage{xcolor,colortbl}

\usepackage{graphicx}
\usepackage{psfrag}
\usepackage[nohead, top=2.9cm, bottom=3.2cm, left=2.9cm, right=3.1cm]{geometry}

\newtheorem{theorem}{Theorem}[section]

\newtheorem{definition}{Definition}[section]

\newtheorem{lemma}{Lemma}[section]

\def\q5uad{\quad\quad\quad\quad\quad}
\def\mytab{\phantom{xxx}}

\title{Encoding Watermark Numbers as Reducible Permutation Graphs
using Self-inverting Permutations \vspace{0.2cm}}

\author{Maria~Chroni \ \ Stavros~D.~Nikolopoulos \ \ Leonidas~Palios}

\date{}

\begin{document}

\maketitle

\vspace{-0.5cm}

\centerline{\it Department of Computer Science \& Engineering}

\centerline{\it University of Ioannina}

\centerline{\it GR-45110 \ Ioannina, Greece}

\centerline{\tt \{mchroni,stavros,palios\}@cs.uoi.gr}


\vskip 0.3in

\begin{abstract}
Several graph theoretic watermark methods have been proposed
to encode numbers as graph structures in software watermarking environments.
In this paper we propose an efficient and easily implementable codec system
for encoding watermark numbers as reducible permutation flow-graphs
and, thus, we extend the class of graphs used in such a watermarking environment.
More precisely, we present
an algorithm for encoding a watermark number~$w$ as
a self-inverting permutation~$\pi^*$,
an algorithm for encoding the self-inverting permutation~$\pi^*$
into a reducible permutation graph~$F[\pi^*]$
whose structure resembles the structure of real program graphs,
as well as decoding algorithms which extract
the permutation~$\pi^*$ from the reducible permutation graph~$F[\pi^*]$
and the number~$w$ from $\pi^*$.
Both the encoding and the decoding process takes time and space linear
in the length of the binary representation of $w$.
The two main components of our proposed codec system,
i.e., the self-inverting permutation $\pi^*$ and
the reducible permutation graph~$F[\pi^*]$, incorporate
the binary representation of the watermark~$w$ in their structure
and possess important structural properties,
which make our system resilient to attacks;
to this end, we experimentally evaluated our system under edge modification attacks
on the graph~$F[\pi^*]$ and the results show that we can detect such attacks
with high probability.

\vspace*{0.1in}
\noindent
\textbf{Keywords:} \ Watermarking, self-inverting permutations,
reducible permutation graphs, polynomial codec algorithms,
structural properties, evaluation.
\end{abstract}

\vspace*{0.05in}
\section{Introduction}
\label{sec:Introduction}

Software watermarking is a technique that is currently being studied
to prevent or discourage software piracy and copyright infringement.
The idea is similar to digital (or media) watermarking where
a unique identifier is embedded in image, audio, or video data
through the introduction of errors not detectable by human perception
\cite{CKLS96}.
The \emph{software watermarking problem} can be described as
the problem of embedding a structure $w$ into a program $P$
producing a program~$P_w$ such that $w$ can be reliably located and
extracted from $P_w$ even after $P_w$ has been subjected to
code transformations such as translation, optimization, and
obfuscation \cite{MC06}.
More precisely, given a program~$P$, a watermark~$w$, and a key~$k$,
the software watermarking problem can be formally described by
the following two functions:
{\tt embed}$(P, w, k)$ $\rightarrow$ $P_w$ and
{\tt extract}$(P_w, k)$ $\rightarrow$ $w$.

Although digital watermarking has made considerable progress and
become a popular technique for copyright protection of
multimedia information \cite{CKLS96,PSSB16,TNMM04}, research on
software watermarking has only recently received considerable attention.
The patent by Davidson and Myhrvold \cite{DM96} presented
the first published software watermarking algorithm.
The preliminary concepts of software watermarking also appeared
in the patents \cite{MC96,S94}.
Collberg et al. \cite{CT99} presented detailed definitions
for software watermarking.
Zhang et al.~\cite{ZYNN03} and Zhu et al.~\cite{ZTW05} have given
brief surveys of software watermarking research (see also Collberg and
Nagra \cite{CN2010} for an exposition of the main results).

\vspace*{0.2in}
\noindent {\bf Graph-based Software Watermarking}. Recently, several graph-based software watermarking techniques have been proposed that encode identification data $w$ as graph structures $G[w]$ and embed them into software ensuring functionality, usability and reversibility.
We refer to the identification data $w$ as the {\it identifier} and to the graph
structure $G[w]$ as the {\it watermark graph}; we may regard identifiers as numbers
(integers in this paper) and refer to them as watermark numbers or, simply, watermarks.

A typical graph-based software watermarking system is mainly comprised of the following four functions:

\vspace*{-0.05in}
\begin{itemize}
\item[$\bullet$\,] {\tt encoder:}
it makes use of an encoding function
{\tt encode} which converts a watermark~$w$ into
a graph~$G[w]$, i.e., ${\tt encode}(w) \rightarrow G[w]$;
\vspace*{-0.05in}
\item[$\bullet$\,] {\tt embedder:} it mainly uses a function which takes as input the
program $P$ (either binary or source code), the intended watermark graph $G[w]$,
and possible a secret key~$k$, and returns the modified program $P_w$ containing the
 graph $G[w]$, i.e., ${\tt embed}(P, G[w], k) \rightarrow P_w$;
\vspace*{-0.05in}
\item[$\bullet$\,] {\tt extractor:} it undertakes to retrieve the watermark graph $G[w]$
from the watermarked program $P_w$ using an appropriate function, i.e., ${\tt extract}(P_w) \rightarrow G_w$;
\vspace*{-0.05in}
\item[$\bullet$\,] {\tt decoder:} it consists of a decoding function {\tt decode}
which converts the watermark graph $G[w]$
into the watermark $w$, i.e., ${\tt decode}(G[w]) \rightarrow w$.
\end{itemize}

\vspace*{-0.05in}
\noindent In this domain, we usually call the pair $({\tt encode}, {\tt decode})_{G[w]}$
\emph{codec system} and refer to both functions {\tt encode} and {\tt decode} as
\emph{codec algorithms}\cite{CKCT03}. In a similar manner, we may use the terms \emph{embex system}
and \emph{embex algorithms} for the pair $({\tt embed}, {\tt extract})_{G[w]}$ and the corresponding {\tt embed} and {\tt extract} functions, respectively.

\begin{itemize}
\item[{\bf (I)}\,] {\bf Codec systems}. While designing a codec system $({\tt encode}, {\tt decode})_\mathcal{G}$ that is appropriate for use
in a real software watermarking environment, we are mainly looking for
a class of graphs $\mathcal{G}$, along with the corresponding functions {\tt encode} and {\tt decode}, having
the following desirable properties and characteristics:

\vspace*{-0.05in}
\begin{itemize}
\item[$\circ$\,]
{\it appropriate graph types}: graphs in $\mathcal{G}$ should be directed having
appropriate properties (e.g., nodes with small outdegree)
so that their structure resembles that of real program graphs;
%
\item[$\circ$\,]
{\it high resilience}: the function ${\tt decode}(G[w])$ should be insensitive to small changes
of $G[\cdot]$ (e.g., insertions or deletions of a constant number of nodes
or edges), that is, if $G[w] \in \mathcal{G}$ and
${\tt decode}(G[w]) \rightarrow w$ then
${\tt decode}(G'[w]) \rightarrow w$ with $G'[w] \approx G[w]$;
%
\item[$\circ$\,]
{\it small size}: the size $|P_w|-|P|$ of the embedded watermark graph $G[w]$ should be small;
%
\item[$\circ$\,]
{\it efficient codecs}: both {\tt encode} and {\tt decode} functions should be polynomially computable.
\end{itemize}

\vspace*{-0.03in}
\noindent In this paper, we focus on the codec part of a software watermark system and propose the codec $({\tt encode}, {\tt decode})_{F[\pi^*]}$ which
incorporates several of the above properties and characteristics making it appropriate for practical use.

\item[{\bf (II)}\,] {\bf Embex systems}. On the other hand, for the design of an efficient embex system $({\tt embed}, {\tt extract})_\mathcal{G}$, we are usually looking for techniques which associate a program $P$ to a directed graph $G[P]$ representing the structure of $P$ as sequences of instructions and methods which handle the graph $G[P]$ and the watermark graph $G[w]$ in an appropriate way; that is the reason we require the structure of $G[w]$ produced by a codec system resembling that of real program graphs. Such a graph $G[P]$ may be the {\it control flow-graph} (CFG) of $P$ which can be obtained by means of a static analysis \cite{NNH04}. In a straight-forward approach, the {\tt embedder} inserts appropriate code into $P$, thus producing the watermarked program $P_w$, so that the watermark graph $G[w]$ shows up as an induced subgraph of $G[P_w]$. In turn, the {\tt extractor} retrieves that subgraph $G[w]$ of $P_w$ and passes it to the codec system, where its {\tt decoder} converts the watermark graph $G[w]$ into $w$. Note that, the embedding process must preserve program semantics, that is, $P$ and $P_w$ must have the same behavior.

\end{itemize}

\noindent We should mention that a software watermarking system usually contains of another function namely {\tt recognizer}: it takes the program $P_w$, the watermark $w$ and the key $k$ as input and returns how confident we are that the $P_w$ contains $w$, i.e., {\tt recognize}$(P_w, w, k)$ $\rightarrow$ $[0.0, 1.0]$ \cite{CN2010}.

\vspace*{0.2in}
\noindent {\bf Techniques and Previous Results}.
The major software watermarking algorithms currently available are based on a number of techniques, such as the register allocation,
spread-spectrum, opaque predicate, abstract interpretation, and dynamic path techniques (see \cite{A02,CCDHKLS04,CC04,CHC03,MIMIT00,NT04}).

In general, according to Collberg and Thomborson's informal taxonomy \cite{CT99}, the software watermarking techniques can be broadly
divided into two main categories, namely, \emph{static} and \emph{dynamic}: in a static technique the watermark $w$ is stored inside the program $P$ in a certain format, either as data or code, and its extraction from the watermarked program $P_w$ requires no execution of $P_w$, whereas in a dynamic one $w$ is stored in $P$ during the execution stage, perhaps only after a particular sequence of input has been used, and it might be retrieved by analyzing the data structures built when $P_w$ is running; see also \cite{CN2010,DM96,MC96,VVS01}. We should also point out that a different software watermarking technique, namely, \emph{abstract} watermarking, has also been proposed: in an abstract framework the watermark $w$ is built in memory, in an abstract data structure, only when $P$ is executed on a particular abstract domain and its extraction requires static analysis on $P_w$ using some abstract interpretation of the semantics of $P_w$ \cite{CC04,GMP17,G08}.

We next report some of the pioneering results in the area of graph-based software watermarking. Indeed, in 1996 Davidson and Myhrvold \cite{DM96} proposed
the first static algorithm which embeds the watermark by reordering the basic blocks of
a control flow-graph. Based on this idea, Venkatesan, Vazirani, and Sinha \cite{VVS01}
proposed the first graph-based software watermarking algorithm
which embeds the watermark by extending a method's control flow-graph
through the insertion of a directed subgraph; it is a static algorithm and is called {\tt VVS} or {\tt GTW}.
Collberg et al. \cite{CHCTS09} proposed the first publicly available implementation ({\tt GTW$_{\tt sm}$}) of algorithm {\tt GTW}; in {\tt GTW$_{\tt sm}$} the watermark is encoded as a reducible
permutation graph (RPG) \cite{CKCT03}, which is a reducible control
flow-graph with maximum out-degree of two, mimicking real code.
The first dynamic watermarking algorithm ({\tt CT}) was proposed by
Collberg and Thomborson \cite{CT99}; it embeds the watermark through
a graph structure which is built on a heap at runtime.
Recently, authors of this paper have contributed in this area
by proposing several codec and embex systems \cite{CN10,CN12,CCN13,MN16}.

\vspace*{0.2in}
\noindent {\bf Attacks}.
A successful attack against the watermarked program~$P_w$ prevents
the recognizer from extracting the watermark while
not seriously harming the performance or correctness of
$P_w$. It is generally assumed that the attacker
has access to the algorithm used by the embedder and recognizer.
There are four main ways to attack a watermark $w \equiv G[w]$ stored in $P_w$:

\vspace*{-0.05in}
\begin{itemize}
\item[$\circ$\,] {\it additive attacks}:
encode a new watermark $w'$ and embed the corresponding watermark graph $G[w']$ into software $P$,
so that an ambiguity is caused and thus the original copyright owners of the software cannot prove
their ownership;
\vspace*{-0.05in}
\item[$\circ$\,] {\it subtractive attacks}:
remove the watermark $G[w]$ of the watermarked software $P_w$
without affecting the functionality of the watermarked software;
\vspace*{-0.05in}
\item[$\circ$\,] {\it distortive attacks}:
modify the watermark graph $G[w]$ to prevent it from being extracted by
the copyright owners and still keep the usability of the software (in this case the {\tt decoder} fails to return any output);
\vspace*{-0.05in}
\item[$\circ$\,] {\it recognition attacks}:
modify the watermark $G[w]$ so that the recognizer gives a misleading result, that is, the {\tt extractor} retrieves the graph $G[w']$ and the {\tt decoder} returns $w' \neq w$.
\end{itemize}

\noindent Typical attacks against the watermark graph $G[w]$ can mainly occur in the following three ways:
$(i)$ {edge-flip attacks}, $(ii)$ {edge-addition/deletion attacks}, and $(iii)$ {node-addition/deletion attacks}.

\vspace*{0.2in}
\noindent {\bf Our Contribution}.
In this paper, we present an efficient and easily implementable codec system for encoding integer numbers as
reducible permutation graphs, whose structure resembles that of real program graphs, through
the use of self-inverting permutations (or SiP, for short).

More precisely, we first present an efficient algorithm which encodes an integer watermark number
 $w$ as a self-inverting permutation~$\pi^*$. Our algorithm, which we call {\tt Encode\_W.to.SiP},
takes as input an integer~$w$, computes its binary representation, constructs
a bitonic permutation on $n^*=2n+1$ numbers, and finally produces a self-inverting permutation~$\pi^*$
of length $n^*$ in $O(n^*)$ time and space. We also present the corresponding decoding algorithm
{\tt Decode\_SiP.to.W}, which converts the permutation $\pi^*$ into the integer~$w$ within
the same time and space complexity.

Having designed an efficient method for encoding integers as self-inverting permutations, we next describe
an algorithm for encoding a self-inverting permutation~$\pi^*$ of length $n^*$ as a reducible permutation graph or, equivalently, watermark flow-graph $F[\pi^*]$. In particular, we propose the algorithm {\tt Encode\_SiP.to.RPG} which exploits
domination relations on the elements of $\pi^*$ and properties of a DAG representation of
$\pi^*$, and produces a reducible permutation flow-graph $F[\pi^*]$ on $n^*+2$ nodes; the whole encoding
process takes $O(n^*)$ time and requires $O(n^*)$ space.
The corresponding decoding algorithm {\tt Decode\_RPG.to.SiP}
extracts the self-inverting permutation~$\pi^*$ from the graph~$F[\pi^*]$ by first converting
it into a directed tree~$T_d[\pi^*]$ and then applying DFS-search on $T_d[\pi^*]$.
The decoding process takes time and space linear in the size of the flow-graph~$F[\pi^*]$, that is, the decoding algorithm takes
$O(n^*)$ time and space; recall that the length of the permutation~$\pi^*$ and the size of the flow-graph~$F[\pi^*]$
are both $O(n^*)=O(n)$, where $n=\lceil \log_2 w \rceil$.

Our codec algorithms are very simple, use elementary operations on
sequences and linked structures, and have very low time and space complexity.
Moreover, both the permutation~$\pi^*$ and the flow-graph~$F[\pi^*]$
incorporate the binary representation of the watermark~$w$ in their structure and
thus possess important structural properties; experimental evaluation results
substantiate that edge modifications attackes on the graph $F[\pi^*]$ can be detected
with high probability.

\vspace*{0.2in}
\noindent {\bf Road Map}.
The paper is organized as follows:
In Section~\ref{sec:Theoretical-Framework} we establish the notation and
related terminology and present background results.
In Section~\ref{sec:System-Components} we describe
the main components of our codec system and present the encoding and decoding
algorithms for the two main phases of our system, namely W-SiP and SiP-RPG.
In Sections~\ref{sec:Properties-SiP} and \ref{sec:Properties-RPG}
we provide structural properties and characterizations of
the self-inverting permutation~$\pi^*$ and
the reducible permutation graph~$F[\pi^*]$, while in
Section~\ref{sec:Detecting-Attacks} we experimentally show that these properties
help prevent edge and/or node modifications attacks.
Finally, in Section~\ref{sec:Concluding-Remarks} we conclude
the paper and discuss possible future extensions.

\vspace*{0.1in}
\section{Theoretical Framework}
\label{sec:Theoretical-Framework}

\noindent
In this section, we present background results and
the main components, namely, the self-inverting permutations (SiP)
and the reducible permutation graphs (RPG), which are used
in the design of our codec system.

\vspace*{0.1in}
\subsection{Preliminaries}
\label{subsec:Preliminaries}

We consider finite graphs with no multiple edges.
For a graph~$G$, we denote by $V(G)$ and $E(G)$
the vertex set and edge set of $G$, respectively.
The \emph{neighborhood}~$N(x)$ of a vertex~$x$ of the graph~$G$ is
the set of all the vertices of $G$ which are adjacent to $x$. The
\emph{degree} of a vertex~$x$ in the graph~$G$, denoted $deg(x)$, is
the number of edges incident on $x$; thus, $deg(x) = |N(x)|$.
For a node~$x$ of a directed graph $G$, the number of directed edges
coming in $x$ is called the \emph{indegree} of $x$
and the number of directed edges leaving $x$ is its \emph{outdegree}.



Next, we introduce some definitions that are key to
our algorithms for encoding numbers as graphs.
Let $\pi$ be a permutation
over the set $N_n = \{1, 2, \ldots, n\}$. We think of permutation $\pi$
as a sequence $(\pi_1, \pi_2, \ldots, \pi_n)$, so, for example, the
permutation $\pi = (1, 4, 2, 7, 5, 3, 6)$ has $\pi_1 = 1$, $\pi_2 = 4$, etc.
By $\pi^{-1}_i$ we denote the position in the sequence
of number $i \in N_n$; in our example, $\pi_4^{-1} = 2$,
$\pi_7^{-1} = 4$, $\pi_3^{-1} = 6$, etc \cite{Gol80}.
The \emph{length} of a permutation $\pi$ is the number
of elements in $\pi$.
The \emph{reverse} of $\pi$, denoted $\pi^R$,
is the permutation $\pi^R = (\pi_n, \pi_{n-1}, \ldots, \pi_1)$.
The \emph{inverse} of $\pi$ is the permutation
$\tau=(\tau_1, \tau_2, \ldots, \tau_n)$ with
$\tau_{\pi_i} = \pi_{\tau_i} = i$. For example, the inverse of the permutation
$\pi = (2,5,1,4,3)$ is the permutation $\tau = (3,1,5,4,2)$.
Clearly, every permutation has a unique inverse, and
the inverse of the inverse is the original permutation.

A \emph{subsequence} of a permutation $\pi=(\pi_1, \pi_2, \ldots, \pi_n)$ is a sequence
$\sigma=(\pi_{i_1}, \pi_{i_2}, \ldots ,\pi_{i_k})$ such that
$i_1 < i_2 < \cdots < i_k$.
If, in addition, $\pi_{i_1} < \pi_{i_2} < \cdots < \pi_{i_k}$, then
we say that $\sigma$ is an \emph{increasing subsequence} of $\pi$,
while if $\pi_{i_1} > \pi_{i_2} > \cdots > \pi_{i_k}$ we say that
$\sigma$ is a \emph{decreasing subsequence} of $\pi$; the length $|\sigma|$ of
a subsequence $\sigma$ is the number of elements in $\sigma$.

The {\it concatenation}
$\sigma_1 \,||\, \sigma_2 \,|| \cdots ||\, \sigma_k$
of $k$ subsequences $\sigma_1, \sigma_2, \ldots, \sigma_k$ of
a permutation $\pi$ is a sequence $\sigma$ of length
$|\sigma_1|+|\sigma_2|+\cdots+|\sigma_k|$ such that
for $1 \leq j \leq k$ and $1 \le i \le |\sigma_j|$,
the $(|\sigma_1|+|\sigma_2|+\cdots+|\sigma_{j-1}|+i)$th element
of $\sigma$ is equal to the $i$th element of $\sigma_j$
where, by convention, $|\sigma_0|=0$.
Additionally,
we denote by $\pi \backslash \sigma$
the subsequence which results from $\pi$ after having ignored
the elements of the subsequence~$\sigma$;
more generally, we denote by $\pi \backslash \{\sigma_1, \sigma_2, \ldots, \sigma_k\}$
the subsequence which results from $\pi$ after having ignored
the elements of the subsequences $\sigma_1$, $\sigma_2$, \ldots, $\sigma_k$.

A \emph{cycle} of a permutation $\pi=(\pi_1, \pi_2, \ldots, \pi_n)$
is an index sequence $c=(i_1, i_2, \ldots, i_p)$ with
$\pi_{i_1}=i_2$, $\pi_{i_2}=i_3$, $\cdots$, $\pi_{i_p}=i_1$.
For example, the permutation $\pi = (4, 7, 1, 6, 5, 3, 2)$
has three cycles $c_1=(1,4,6,3)$, $c_2=(2,7)$, and $c_3=(5)$ of
lengths 4, 2, and 1, respectively. In general, a permutation~$\pi$
contains $\ell$ cycles, where $1 \leq \ell \leq n$; for example,
the identity permutation over the set $N_n$ contains
$n$ cycles of length 1. Throughout the paper, a cycle of length~$k$
is referred to as a $k$-cycle.

A \emph{left-to-right maximum} (\emph{left-to-right minimum}, resp.)
of $\pi$ is an element~$\pi_i$, $1 \leq i \leq n$, such that
$\pi_j < \pi_i$ ($\pi_j > \pi_i$, resp.) for all $j<i$.
The increasing (decreasing, resp.) subsequence
$\sigma=(\pi_{i_1}, \pi_{i_2}, \ldots ,\pi_{i_k})$ is
a \emph{left-to-right maxima (minima, resp.) subsequence} if
it consists of all the left-to-right maxima (minima, resp.) of $\pi$;
clearly, $\pi_{i_1}=\pi_1$.
For example, the left-to-right maxima subsequence of the permutation
$\pi = (5,6,2,8,1,9,7,4,3)$ is $(5,6,8,9)$,
while the left-to-right minima subsequence of $\pi$ is $(5,2,1)$.

The \emph{$1$st increasing (decreasing, resp.) subsequence} $S_1$
of a permutation $\pi$ is defined to be the left-to-right maxima
(minima, resp.) subsequence of $\pi$.
The \emph{$i$th increasing (decreasing, resp.) subsequence} $S_i$
of $\pi$ is defined to be the left-to-right maxima (minima, resp.)
subsequence of $\pi'$, where $\pi'$ results from $\pi$ after
having ignored the elements of the $1$st, $2$nd, \ldots, $(i-1)$st
increasing (decreasing, resp.) subsequences of $\pi$, i.e.,
$\pi'=\pi \backslash \{S_1, S_2, \ldots, S_{i-1}\}$.
For example, the increasing subsequences of the permutation
$\pi = (5,6,2,8,1,9,7,4,3)$ are
$S_1 = (5, 6, 8, 9)$, $S_2 = (2, 7)$ since $\pi'=\pi \backslash S_1 = (2, 1, 7, 4, 3)$,
$S_3 = (1, 4)$, and $S_4 = (3)$, while its decreasing subsequences are
$S_1 = (5, 2, 1)$, $S_2 = (6, 4, 3)$, $S_3 = (8, 7)$, and
$S_4 = (9)$.

We say that an element~$i$ of a permutation~$\pi$ over the set~$N_n$
\emph{dominates} the element $j$ if $i > j$ and
$\pi^{-1}_i < \pi^{-1}_j$.
An element~$i$ \emph{directly dominates} (or d-dominates, for short)
the element~$j$ if $i$ dominates $j$ and there exists no element~$k$
in $\pi$ such that $i$ dominates $k$ and $k$ dominates $j$;
for example, in the permutation $\pi = (8, 3, 2, 7, 1, 9, 6, 5, 4)$,
the element~$7$ dominates the elements $1, 6, 5, 4$ and
directly dominates the elements $1, 6$.
Let $d$-$dom(j)$ be the set of all the elements of
a permutation~$\pi$ which d-dominate the element~$j$ and
$dmax(j)$ be the element of the set~$d$-$dom(j)$ with maximum value;
for example, in $\pi = (8, 3, 2, 7, 1, 9, 6, 5, 4)$,
$d$-$dom(6)=(7,9)$ and $dmax(6)=9$,
and $d$-$dom(1)=(2,7)$ and $dmax(1)=7$.
By definition, the element~$i$ of a permutation $\pi$
such that $i=dmax(j)$ is the rightmost element on the left of $j$
in $\pi$ that d-dominates $j$.

\vspace*{0.1in}
\subsection{Self-inverting Permutations (SiP)}

\vspace*{0.1in}
\noindent
We next define the main component of our codec system,
namely, the self-inverting permutation (SiP), and
prove key properties for encoding numbers as reducible permutation graphs.

\begin{definition}  \label{def:sip}  
Let $\pi=(\pi_1, \pi_2, \ldots, \pi_n)$ be a permutation over
the set $N_n$.
A \emph{self-inverting permutation} (or involution) is
a permutation that is its own inverse: $\pi_{\pi_i} = i$.
\end{definition}

The definition of the inverse of a permutation implies that
a permutation is a self-inverting permutation iff
all its cycles are of length 1 or 2;
hereafter, we shall denote a 2-cycle by $(x,y)$ with $x > y$
and a 1-cycle by $(x)$ or, equivalently, $(x, x)$.

\begin{definition}  \label{def:cycle_representation}  
A sequence $C = (c_1, c_2, \ldots, c_k)$ of all the 2- and 1-cycles
of a self-inverting permutation~$\pi$ is
a \emph{decreasing cycle representation} of $\pi$
if $c_1 \succ c_2 \succ \cdots \succ c_k$ where
$c_i = (a_i,b_i) \succ c_j = (a_j,b_j)$
(with $a_i \geq b_i$ and $a_j \geq b_j$)
if $b_i > b_j$, $1 \leq i, j \leq k$.
The cycle~$c_k$ containing the smallest element among the elements
of the cycles is the \emph{minimum element} of the sequence~$C$.
\end{definition}

\vspace*{0.1in}
\subsection{Reducible Permutation Graphs (RPG)}

\vspace*{0.1in}
\noindent
A directed graph~$G$ is \emph{strongly connected} if for every ordered pair
of vertices $(x, y)$ of $G$ there is a directed path in $G$ from $x$
to $y$. A node~$y$ is an \emph{entry} for a subgraph~$H$ of the graph~$G$ if
there is an edge $(x,y)$ in $G$ such that $y \in H$ and $x \not\in H$.

\begin{definition}  \label{def:reducible}  
A \emph{flow-graph} $G$ is a directed graph with a source node $s \in V(G)$
from which all other nodes are reachable. A flow-graph is \emph{reducible} if it does not have
a strongly connected subgraph with two (or more) entries.
\end{definition}

We can equivalently define a reducible flow-graph $G$ as a directed graph with a source node $s \in V(G)$ such that
every node of $G$ is reached from $s$ and every directed path from $s$ to a directed cycle $C$ reaches $C$ at the same node.

There are at least two other equivalent definitions, as
Theorem~\ref{thm:thmA} shows.
These definitions use a few more graph-theoretic concepts.
For some node $x$, the edge $(x, x)$ is a \emph{cycle-edge}.
A depth first search (DFS) traversal of a graph $G$ partitions
its edges into \emph{tree}, \emph{forward}, \emph{back}, and \emph{cross} edges. The tree, forward, and cross edges
of $G$ form a dag known as a \emph{DFS dag}.

\begin{theorem}   \label{thm:thmA}   
\cite{HU72,HU74} Let $F$ be a flow-graph.
The following three statements about $F$ are equivalent:
\begin{itemize}
\vspace*{0.0in}
\item[(i)\,]
the graph~$F$ is reducible;
\vspace*{0.0in}
\item[(ii)\,]
the graph~$F$ has a unique DFS dag;
\vspace*{0.0in}
\item[(iii)\,]
the graph~$F$ can be transformed into a single node
by repeated application of the transformations ${\phi}_1$
and ${\phi}_2$,
where ${\phi}_1$ removes a cycle-edge, and ${\phi}_2$
picks a non-initial node~$y$ that has only one
incoming edge~$(x,y)$ and glues nodes $x$ and $y$.
\end{itemize}
\end{theorem}

\noindent It is well-known that a reducible flow-graph has at most one Hamiltonian path \cite{CKCT03}.

\vspace*{0.1in}
\section{Our Codec System}
\label{sec:System-Components}

\begin{figure}[t]
    \hrule\medskip\medskip
    \centering
    \includegraphics[scale=0.55]{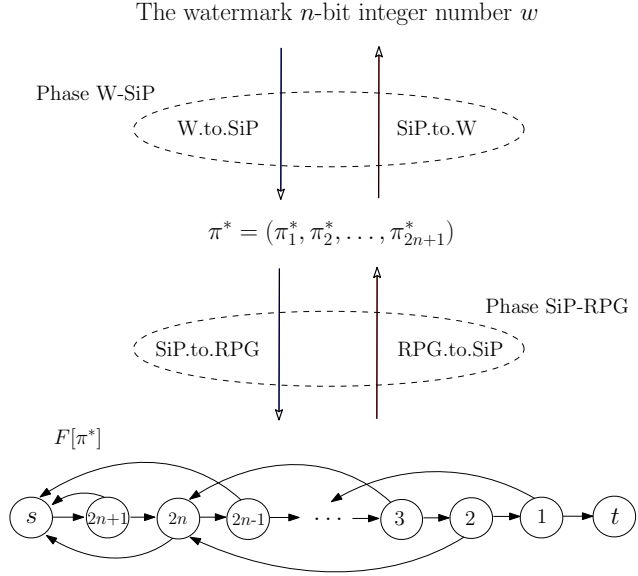}
    \centering
    \medskip\medskip\hrule\medskip
    \caption{\small{The main data components used by
the codec system W-RPG for encoding a watermark number~$w$:
(i) the self-inverting permutation~$\pi^*$ and
(ii) the reducible permutation graph~$F[\pi^*]$.}}
\medskip\medskip
\label{fig:fig1}
\end{figure}

For encoding a watermark number~$w$, our codec system uses two main components:
(i)~the self-inverting permutation~$\pi^*$ and
(ii)~the reducible permutation graph~$F[\pi^*]$; see Figure~\ref{fig:fig1}.
The same figure also depicts the two main phases of our codec system process:

\vspace*{0.0in}
\begin{itemize}
\item[(I)] {\bf Phase W--SiP:} it uses two algorithms,
one for encoding the watermark number~$w$ into
a self-inverting permutation~$\pi^*$ and
the other for extracting $w$ from $\pi^*$;

\vspace*{0.02in}
\item[(II)] {\bf Phase SiP--RPG:} this phase uses two algorithms
as well, one for encoding the self-inverting permutation~$\pi^*$
into a reducible permutation graph~$F[\pi^*]$ and
the other for extracting $\pi^*$ from $F[\pi^*]$.
\end{itemize}

\noindent
Our codec system encodes an integer~$w$ as
a self-inverting permutation~$\pi^*$ using a construction technique
which captures into $\pi^*$ important structural properties (see Section~\ref{sec:Properties-SiP}).
As we shall see in Section~\ref{sec:Detecting-Attacks},
these properties enable an attack-detection system to identify
edge and/or node modifications made by an attacker to $\pi^*$.
Moreover, the encoding approach adopted in our system enables it
to encode any integer~$w$ as a self-inverting permutation~$\pi^*$
of length $n^*=2n+1$, where $n=2 \lceil \log_2 w \rceil + 1$.

\vspace*{0.05in}
\noindent
The reducible permutation graph~$F[\pi^*]$ produced by
our system's algorithms consists of $n^*+2$ nodes, say,
$u_{n^*+1}, u_{n^*}, \ldots, u_i, \ldots, u_0$,
which include:

\begin{enumerate}
\item[(A)] {\bf A header node}:
it is a root node with outdegree 1 from which all other nodes
of the graph~$F[\pi^*]$ are reachable; note that every control
flow-graph has such a node.
In $F[\pi^*]$ the header node is denoted by $s=u_{n^*+1}$;

\vspace*{0.01in}
\item[(B)] {\bf A footer node}:
it is a node with outdegree 0 that is reachable from all other
nodes of the graph~$F[\pi^*]$. Every control flow-graph has
such a node representing the exit of the method.
In $F[\pi^*]$ the footer node is denoted by $t=u_{0}$;

\vspace*{0.01in}
\item[(C)] {\bf The body}:
it consists of $n^*$ nodes $u_{n^*}, u_{n^*-1}, \ldots, u_i, \ldots, u_1$
each with outdegree 2.
In particular, each node $u_i$ $(1 \leq i \leq n^*)$ has exactly
two outgoing pointers: one points to node $u_{i-1}$ and
the other points to a node~$u_m$ with $m > i$;
recall that $u_{n^*+1}=s$ and  $u_{0}=t$.
\end{enumerate}

\vspace*{0.1in}
\noindent
By construction, the reducible permutation graph~$F[\pi^*]$
is of order (i.e., number of nodes) $n^*+2$ and
size (i.e., number of edges) $2n^*+1$.
Thus, since $n^*=2n+1$, both the order and size of
graph~$F[\pi^*]$ are of $O(n)$, where
$n=2 \lceil \log_2 w \rceil + 1$.

Recall that our contribution in this paper has to do with
both the W--SiP and the SiP--RPG phase.
We design and analyze algorithms for encoding a watermark
number~$w$ as a SiP~$\pi^*$ and algorithms for encoding a SiP~$\pi^*$
as a reducible permutation flow-graph~$F[\pi^*]$ along with
the corresponding decoding algorithms; we also show properties of
our codec system that prevent edge and/or node modification attacks.

\vspace*{0.2in}
\subsection{Codec Algorithms for Phase W-SiP}
\label{subsec:W-to-SiP}

In this section, we first introduce the notion of
a \emph{Bitonic Permutation} and then we present two algorithms,
namely {\tt Encode\_W.to.SiP} and {\tt Decode\_SiP.to.W},
for encoding an integer~$w$ into a self-inverting permutation~$\pi^*$
and for extracting it from $\pi^*$, respectively.
Both algorithms run in $O(n)$ time, where $n = \log_2 w$ is
the length of the binary representation of the integer~$w$ \cite{CN10}.

\vspace*{0.1in}
\noindent {\bf Bitonic Permutations}:
A permutation $\pi = (\pi_1, \pi_2, \ldots , \pi_n)$ over
the set $N_n$ is called \emph{bitonic} if
it either monotonically increases and then monotonically decreases,
or monotonically decreases and then monotonically increases.
For example, the permutations $\pi_1 = (1, 4, 6, 7, 5, 3, 2)$ and
$\pi_2 = (6, 4, 3, 1, 2, 5, 7)$ are both bitonic. Trivially, an increasing or decreasing permutation is considered bitonic.

Let $\pi = (\pi_1, \pi_2, \ldots, \pi_i, \pi_{i+1}, \ldots, \pi_n)$
be a bitonic permutation over the set $N_n$ that
first monotonically increases and then monotonically decreases and
let $\pi_i$ be the \emph{leftmost} element of $\pi$
such that $\pi_i > \pi_{i+1}$; note that $\pi_i$ is the maximum
element of $\pi$.
Then, we call the sequence $X = (\pi_1, \pi_2, \ldots, \pi_{i-1})$
\emph{the increasing subsequence} of $\pi$ and
the sequence $Y = (\pi_{i}, \pi_{i+1}, \ldots, \pi_n)$
\emph{the decreasing subsequence} of $\pi$. Note that although $(\pi_1, \pi_2, \ldots, \pi_{i})$
is increasing, the increasing subsequence of $\pi$ is defined up to the element $\pi_{i-1}$.

\vspace*{0.1in}
\noindent
{\bf Notations}:
If $B_1 = b_1 b_2 \cdots b_n$ and $B_2 = d_1 d_2 \cdots d_m$
are two binary numbers, then the number $B_1||B_2$ is
the binary number $b_1 b_2 \cdots b_n d_1 d_2 \cdots d_m$.
The \emph{binary sequence} of the number $B = b_1 b_2 \cdots b_n$
is the sequence $B^* = (b_1, b_2, \ldots, b_n)$ of length~$n$.

\begin{figure}[t]
    \hrule\medskip
    \centering
    \includegraphics[scale=0.7]{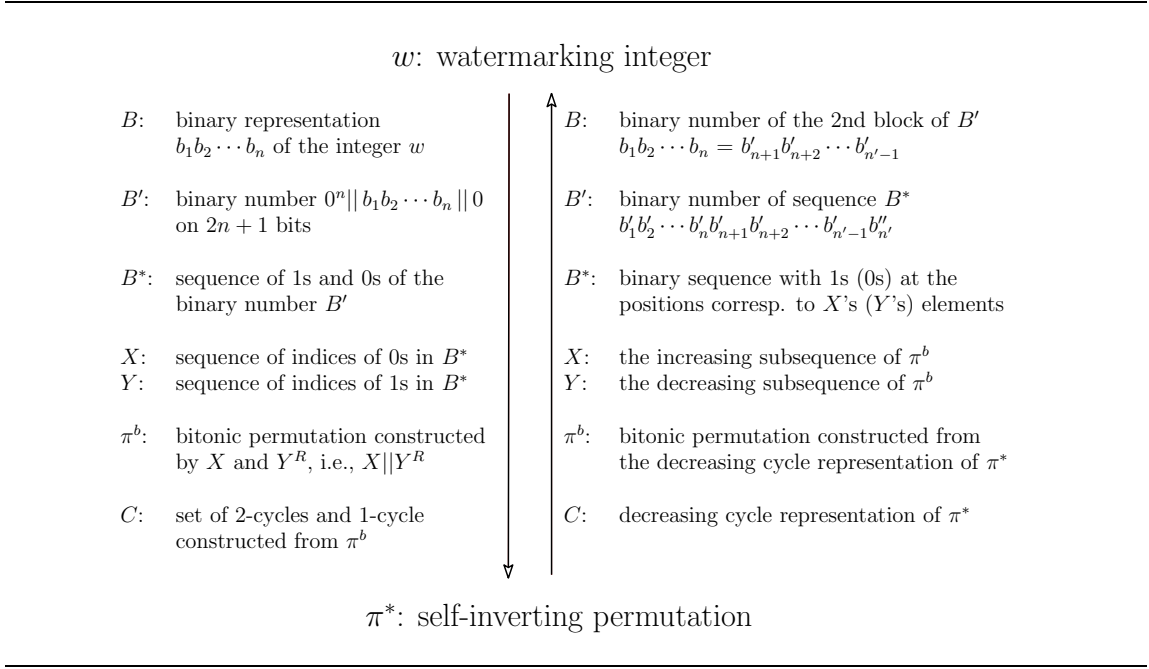}
    \centering
    \medskip\medskip\hrule\medskip
    \caption{\small{The main data components used by
Algorithms {\tt Encode\_W.to.SiP} and {\tt Decode\_SiP.to.W}.}}
\medskip\medskip
\label{fig:fig2}
\end{figure}

\vspace*{0.1in}

\vskip 0.2in 
\subsubsection{Algorithm Encode\_W.to.SiP}
\label{subsubsec:Encode-W-to-SiP}

We next present an algorithm for encoding an integer
as a self-inverting permutation without having to consult a list
of all self-inverting permutations.
Our algorithm takes as input an integer $w$,
computes its binary representation, and then produces
a self-inverting permutation~$\pi^*$ in time linear in
the length of the binary representation of $w$.
The proposed algorithm is the following:

\vspace*{0.15in}
\noindent Algorithm {\tt Encode\_W.to.SiP}
\vspace*{-0.04in}
\begin{enumerate}
\item[1.\,]   
Compute the binary representation $B$ of $w$ and let $n$ be
the length of $B$;
\vspace*{0.0in}
\item[2.\,]   
Construct the binary number $B' = \underbrace{00 \cdots 0}_n||B||0$ of
length $n^* = 2n+1$, and then the binary sequence $B^*$ of $B'$;
\vspace*{0.0in}
\item[3.\,]   
Construct the sequence $X = (x_1, x_2, \ldots, x_k)$ of
the $1$s' positions and the sequence $Y = (y_1, y_2, \ldots, y_m)$ of
the $0$s' positions in $B^*$ from left to right, where $k+m = n^*$;
\vspace*{0.0in}
\item[4.\,]   
Construct the bitonic permutation $\pi^b = X || Y^R =
(x_1, x_2, \ldots, x_k, y_m, y_{m-1}, \ldots, y_1)$ over
the set $N_{n^*} = N_{2n+1}$;
\vspace*{0.0in}
\item[5.\,]   
for $i = 1, 2, \ldots, n = \lfloor n^* / 2 \rfloor$ do\\
\mytab$\{$\textit{construct a 2-cycle with the $i$-th element
of $\pi^b$ from left and the $i$-th element from right}$\}$\\
\mytab construct the 2-cycle
$c_i = (\pi_i^b, \pi_{n^* - i + 1}^b)$;\\
construct the 1-cycle $c_i = (\pi_{n+1}^b)$;
\vspace*{0.0in}
\item[6.\,]   
Initialize the permutation~$\pi^*$ to the identity permutation
$(1, 2, \ldots, 2n+1)$;\\
for each 2-cycle $(\pi_i, \pi_j)$ computed at Step~5,
set $\pi_{\pi_i}^* = \pi_j$ and $\pi_{\pi_j}^* = \pi_i$;
\vspace*{0.0in}
\item[7.\,]   
Return the self-inverting permutation~$\pi^*$;
\end{enumerate}

\vspace*{0.1in}
\noindent
{\bf Example~3.1} Let $w = 12$ be the input watermark integer
in the algorithm {\tt Encode\_W.to.SiP}.
We first compute the binary representation $B = 1100$ of
the number~$12$; then we construct the binary number $B' = 000011000$
and the binary sequence $B^* = (0, 0, 0, 0, 1, 1, 0, 0, 0)$ of
$B'$; we compute the sequences $X = (5, 6)$ and
$Y = (1, 2, 3, 4, 7, 8, 9)$, and then construct
the bitonic permutation
$\pi^b = (5, 6, 9, 8, 7, 4, 3, 2, 1)$ on $n^*=9$ numbers;
since $n^*=9$ is odd, we form four 2-cycles $(5, 1)$, $(6, 2)$, $(9, 3)$, $(8, 4)$ and
one 1-cycle $(7)$, and then construct the self-inverting permutation $\pi^* = (5, 6, 9, 8, 1, 2, 7, 4, 3)$.

\medskip
\noindent
\textit{Time and Space Complexity}.
The encoding algorithm {\tt Encode\_W.to.SiP} performs
basic operations on sequences of $O(n)$ length,
where $n$ is the number of bits in the binary representation
of $w$ (see Figure~\ref{fig:fig2}).
Thus, the whole encoding process requires $O(n)$ time and space,
and the following theorem holds:

\vspace*{0.04in}
\begin{theorem}\label{thm:theo1.1}
Let $w$ be an integer and let $b_1b_2\cdots b_n$ be
the binary representation of $w$.
The algorithm {\tt Encode\_W.to.SiP} encodes the number~$w$
in a self-inverting permutation~$\pi^*$ of length $2n+1$
in $O(n)$ time and space.
\end{theorem}

\vskip 0.2in 
\subsubsection{Algorithm Decode\_SiP.to.W}
\label{subsubsec:Decode-SiP-to-W}

Next, we present an algorithm for decoding a self-inverting permutation. More precisely,
our algorithm, which we call Decode\_SiP.to.W, takes
as input a self-inverting permutation~$\pi^*$ produced by
Algorithm Encode\_W.to.SiP and returns its corresponding integer~$w$.
Its time complexity is linear in
the length of the permutation~$\pi^*$.
We next describe the proposed algorithm:


\vspace*{0.15in} \noindent Algorithm {\tt Decode\_SiP.to.W}
\vspace*{-0.04in}
\begin{enumerate}
\item[1.\,]   
Compute the decreasing cycle representation
$C = (c_1, c_2, \ldots, c_{n+1})$ of the self-inverting permutation $\pi^* = (\pi_1, \pi_2, \ldots, \pi_{n^*})$, where $n^* = 2n+1$;
\vspace*{0.01in}
\item[2.\,]   
Construct the bitonic permutation $\pi^b$ of length $n^*$ as follows:
\vspace*{0.01in}\\
let $C = ((a_1,b_1), (a_2,b_2), \ldots, (a_{n+1},b_{n+1}))$, where $c_i=(a_i,b_i)$ with $a_i > b_i$, $1 \geq i \geq n+1$, and $b_1 > b_2 > \cdots > b_{n+1}$;\\
\mytab $\circ$ \ compute $Q=(a_{n+1}, a_{n}, \ldots, a_{2}, a_{1}, b_{1}, b_{2}, \ldots, b_{n+1})$;\\
\mytab $\circ$ \ delete from $Q$ the element $a_i$ which forms the 1-cycle $(a_i,b_i)$ of $\pi^*$ and thus $a_i=b_i$;\\
\mytab $\circ$ \ set $\pi^b=(\pi^b_1, \pi^b_2, \ldots, \pi^b_{n+1})$ $\leftarrow$ $Q$;
\vspace*{0.01in}
\item[3.\,]   
Construct the increasing subsequence
$X = (\pi^b_1, \pi^b_2, \ldots, \pi^b_k)$ of $\pi^b$ and then
the decreasing subsequence
$Y = (\pi^b_{k+1}, \pi^b_{k+2}, \ldots, \pi^b_{n^*})$, where $\pi^b_{k+1}$ is the top element of $\pi^b$;
\vspace*{0.01in}
\item[4.\,]   
Construct the binary sequence $B^* = (b_1, b_2, \ldots, b_{n^*})$
by setting 1 in positions $\pi^b_1, \pi^b_2, \ldots, \pi^b_k$ and
0 in positions $\pi^b_{k+1}, \pi^b_{k+2}, \ldots, \pi^b_{n^*}$;
\vspace*{0.04in}
\item[5.\,]   
Compute $B' = b'_1 b'_2 \ldots, b'_n, b'_{n+1} \cdots b'_{n^*-1} b'_{n^*}$ from $B^*$;
\vspace*{0.01in}
\item[6.\,]   
Return the decimal value~$w$ of the binary number
$B = b'_{n+1} b'_{n+2} \cdots b'_{n^*-1}$;
\end{enumerate}

\medskip\noindent
The decoding algorithm {\tt Decode\_SiP.to.W} is essentially the reverse of
the corresponding encoding algorithm {\tt Encode\_W.to.SiP}.

\vspace*{0.2in}
\noindent
{\bf Example~3.2}
Let $\pi^* = (5, 6, 9, 8, 1, 2, 7, 4, 3)$ be a self-inverting
permutation produced by Algorithm {\tt Encode\_W.to.SiP}.
The decreasing cycle representation of $\pi^*$ is
the sequence
$C = \bigl( (7,7), \,(8, 4), \,(9, 3), \,(6, 2), \,(5, 1) \bigr)$;
we take the cycles in $C$ from right to left and construct the permutation
$\pi^b = (5, 6, 9, 8, 7, 4, 3, 2, 1)$;
then, we compute the increasing subsequence $X=(5, 6)$ and
the decreasing subsequence $Y = (9, 8, 7, 4, 3, 2, 1)$ of $\pi^b$;
we next construct the binary sequence
$B^* = (0, 0, 0, 0, 1, 1, 0, 0, 0)$ of length $9$ and finally the the sequence $B' = 000011000$ of the elements of $B^*$;
the decimal value of the binary number 1100 is the integer $w = 12$.

\medskip
\noindent
\textit{Time and Space Complexity}.
It is easy to see that the decoding algorithm {\tt Decode\_SiP.to.W}
performs the same basic operations on sequences of $O(n)$ length
as the encoding algorithm (see Figure~\ref{fig:fig2}).
Thus, we obtain the following result:

\vspace*{0.04in}
\begin{theorem}\label{thm:theo1.2}
Let $w$ be an integer (whose binary representation has length~$n$) and
let $\pi^*$ be the self-inverting permutation of length $n^* = 2n+1$
produced by Algorithm {\tt Encode\_W.to.SiP} to encode $w$.
Algorithm {\tt Decode\_SiP.to.W} correctly extracts $w$ from
$\pi^*$ in $O(n^*) = O(n)$ time and space.
\end{theorem}

\vspace*{0.2in}
\subsection{Codec Algorithms for Phase SiP-RPG}
\label{subsec:SiP-to-RPG}

\noindent
In this section, we concentrate on the system's phase SiP--RPG
and present an efficient algorithm for encoding a self-inverting
permutation~$\pi^*$ into a reducible permutation graph~$F[\pi^*]$, along with the corresponding decoding algorithm.

The proposed encoding algorithm, which we call {\tt Encode\_SiP.to.RPG},
takes as input the self-inverting permutation~$\pi^*$ produced by
the algorithm {\tt Encode\_W.to.SiP} and constructs
a reducible permutation flow-graph~$F[\pi^*]$ by using
a DAG representation~$D[\pi^*]$ of the permutation~$\pi^*$;
in fact, it uses a parent-relation of a tree obtained from
the graph~$D[\pi^*]$ defined below.
The whole encoding process takes $O(n^*)$ time and requires
$O(n^*)$ space, where $n^*$ is the length of the input
self-inverting permutation~$\pi^*$.

\begin{figure}[t]
    \hrule\medskip
    \centering
    \includegraphics[scale=0.6]{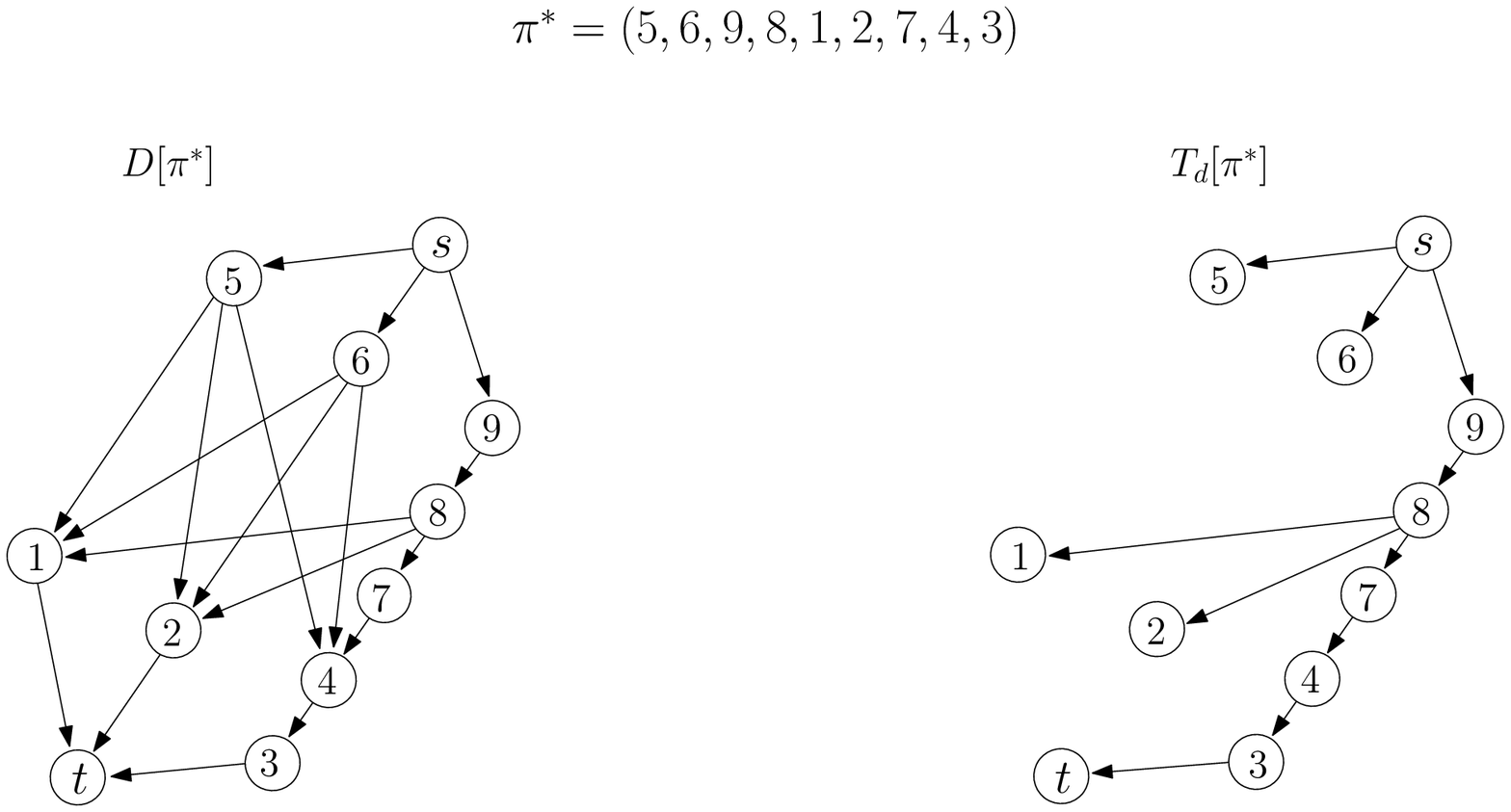}
    \centering
    \medskip\medskip\hrule\medskip
    \caption{\small{The DAG $D[\pi^*]$ of the self-inverting
permutation $\pi^*$ and the corresponding Dmax-tree $T_{d}[\pi^*]$.}}
\medskip\medskip
\label{fig:fig3}
\end{figure}

Next, we first describe the main ideas and the structures behind
our encoding algorithm.
In particular, given a self-inverting permutation~$\pi^*$
we construct a directed acyclic graph and a directed tree
by taking advantage of the $dmax$ values of the elements of $\pi^*$
(recall that $dmax(i)$ with respect to the permutation~$\pi^*$,
where $1 \le i \le n^*$,
is equal to the maximum value element among the elements in $\pi^*$
that d-dominate $i$; see Subsection~\ref{subsec:Preliminaries}).

\vspace*{0.1in}
\noindent {\bf DAG Representation $D[\pi^*]$}:
We construct the directed acyclic graph $D[\pi^*]$ by exploiting
the d-domination relation of the elements of $\pi^*$ as follows:
(i)~for every element $i$ of $\pi^*$, we create a corresponding
vertex~$v_i$ and we add it into the vertex set~$V(D[\pi^*])$
of $D[\pi^*]$;
(ii)~for every pair of vertices $(v_i, v_j)$ where
$v_i, v_j \in V(D[\pi^*])$, we add the directed edge~$(v_i, v_j)$
in $E(D[\pi^*])$ if the element~$i$ d-dominates the element~$j$
in $\pi^*$;
(iii)~we create two dummy vertices $s=v_{n^*+1}$ and $t=v_0$ and
we add them both in $V(D[\pi^*])$; then, we add in $E(D[\pi^*])$
the directed edge~$(s, v_i)$ for every $v_i$ with indegree
equal to $0$, and the edge $(v_j, t)$ for every $v_j$ with
outdegree equal to $0$.

Figure~\ref{fig:fig3} depicts the graph~$D[\pi^*]$ of
the permutation $\pi^* = (5,6,9,8,1,2,7,4,3)$.
Note that, by construction, $i > j$ for every
directed edge~$(v_i, v_j)$ of $D[\pi^*]$ since
the element~$i$ d-dominates the element~$j$ in $\pi^*$.

\vspace*{0.1in}
\noindent {\bf Dmax-tree $T_{d}[\pi^*]$}:
We next construct the directed tree $T_{d}[\pi^*]$,
which we call Dmax-tree,
by exploiting the $dmax$ values of the elements on
the nodes of $D[\pi^*]$.
The Dmax-tree $T_{d}[\pi^*]$ is simply constructed as follows:

\begin{itemize}
\item[(i)\,]
construct the DAG $D[\pi^*]$;

\vspace*{0.00in}
\item[(ii)\,]
delete the directed edge~$(v_i, v_j)$ from $D[\pi^*]$
if $i$ $\neq$ $dmax(j)$.
\end{itemize}

\noindent
The Dmax-tree $T_{d}[\pi^*]$ of the permutation
$\pi^* = (5,6,9,8,1,2,7,4,3)$ is shown in Figure~\ref{fig:fig3}.
We point out that the construction of the Dmax-tree $T_{d}[\pi^*]$
can also be done directly from permutation~$\pi^*$
by computing the element $dmax(i)$ for
each element $i \in \pi^*$, $1 \leq i \leq n^*$;
note that $s=v_{n^*+1}$ dominates all the elements of $\pi^*$.

\vskip 0.3in 
\subsubsection{Algorithm Encode\_SiP.to.RPG}
\label{subsubsec:Encode-SiP-to-RPG}

Given a self-inverting permutation~$\pi^*$ of length $n^*$,
our proposed encoding algorithm {\tt Encode\_W.to.SiP}
works as follows:
first, it computes the $dmax$ value of each of the $n^*$ elements
of the self-inverting permutation $\pi^*$ (Step~1),
which it then uses to construct a directed graph~$F[\pi^*]$
on $n^*+2$ nodes (Steps~2 and 3).
Next, we present the encoding algorithm in detail.


\vspace*{0.15in} \noindent Algorithm {\tt Encode\_SiP.to.RPG}
\vspace*{-0.04in}
\begin{enumerate}
\item[1.\,]
for each element $i \in \pi^*$, $1 \leq i \leq n^*$, do\\
\mytab $\circ$ \ Compute $P(i) = dmax(i)$;
%
\item[2.\,]
Construct a directed graph $F[\pi^*]$ on $n^*+2$ vertices
as follows:\\
\mytab $\circ$ \ $V(F[\pi^*]) = \{s = u_{n^*+1}, u_{n^*}, \ldots,
u_1, u_0 = t\}$;\\
\mytab $\circ$ \ add the forward edges $(u_{i+1}, u_i)$ in $E(F[\pi^*])$, for $0 \leq i \leq n^*$;
%
\item[3.\,]
for each vertex $u_i \in V(F[\pi^*])$, $1 \leq i \leq n^*$, do\\
\mytab $\circ$ \ add the backward edge $(u_i, u_m)$ in $E(F[\pi^*])$, where $m = P(i)$;
%
\item[4.]
Return the graph $F[\pi^*]$;
\end{enumerate}

\vspace*{0.04in}
\noindent
{\it Time and Space Complexity.}
The most time-consuming step of the algorithm is
the computation of the value $dmax(i)$ for each element~$i$
of $\pi^*$ (Step~1). On the other hand, the construction of
the reducible permutation flow-graph $F[\pi^*]$ on $n^*+2$ nodes
requires only the forward edges (Step~2) which can be
trivially computed, and the backward edges (Step~3)
which can be computed using the values of $dmax$.

Returning to Step~1, since $dmax(i)$ is
the rightmost element on the left of the element~$i$
in the permutation~$\pi^*$ that is greater than $i$,
the values~$P(i)$ can be computed
using the input permutation as follows:

\vspace*{-0.02in}
\begin{enumerate}
\item[(i)]
insert the element~$s$ with value~$n^*+1$
into an initially empty stack~$S$;
%
\vspace*{-0.08in}
\item[(ii)]
for each element $\pi_i^*$, $i = 1, 2, \ldots, n^*$,
do the following:\\
\mytab while the element at the top of $S$ is less than
$\pi^*_i$\\
\mytab\mytab pop it from the stack~$S$;\\
\mytab $P(\pi^*_i) =$ element at the top of $S$;\\
\mytab push $\pi^*_i$ into the stack $S$;
\end{enumerate}
\vspace*{-0.02in}

\noindent
For the correctness of this procedure, note that
the contents of the stack~$S$ are in decreasing order
from bottom to top;
in fact, at the completion of the processing of element~$\pi^*_i$,
$S$ contains (from top to bottom) the left-to-right maxima
of the reverse subpermutation
$(\pi^*_i, \pi^*_{i-1}, \ldots, \pi^*_1, n^*+1)$.
Additionally, it is important to observe that the value $n^*+1$
at the bottom of the stack~$S$ is never removed.

The time to process element~$\pi_i^*$ in step~(ii) is $O(1+t_i)$
where $t_i$ is the number of elements popped from the stack~$S$
while processing $\pi_i^*$.
Since the number of pops from $S$ does not exceed the number
of pushes in $S$ and
since each element of the input permutation $\pi^*$
is inserted exactly once in $S$,
the whole computation of the function~$P()$ takes $O(n^*)$ time
and space, where $n^*$ is the length of the permutation~$\pi^*$.
Thus, we obtain the following result.

\vspace*{0.04in}
\begin{theorem}\label{thm:theo5.1}
The algorithm {\tt Encode\_SiP.to.RPG} for encoding
a self-inverting permutation $\pi^*$ of length $n^*$ as
a reducible permutation flow-graph $F[\pi^*]$ requires
$O(n^*)$ time and space.
\end{theorem}

\begin{figure}[t]
    \hrule\medskip
    \centering
    \includegraphics[scale=0.6]{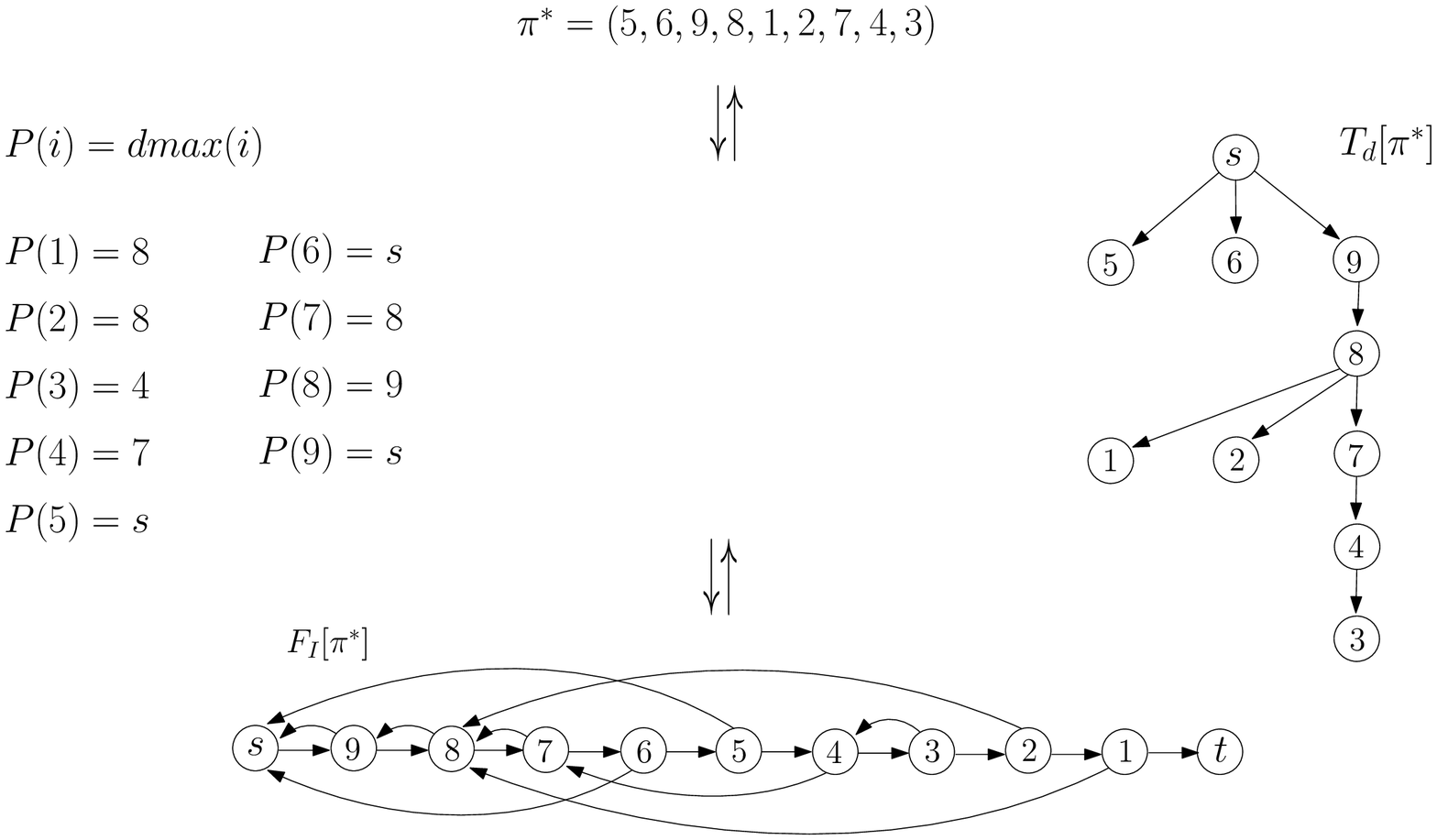}
    \centering
    \medskip\medskip\hrule\medskip
    \caption{\small{The main structures used or constructed
by the algorithms {\tt Encode\_SiP.to.RPG} and
{\tt Decode\_RPG.to.SiP}, that is,
the  self-inverting permutation $\pi^*$,
the function $P()$,
the reducible graph $F_I[\pi^*]=F[\pi^*]$,
and the Dmax-tree $T_{d}[\pi^*]$.}}
\medskip\smallskip
\label{fig:fig4}
\end{figure}

\vskip 0.2in 
\subsubsection{Algorithm Decode\_RPG.to.SiP}
\label{subsubsec:Decode-RPG-to-SiP}

Having presented the encoding algorithm {\tt Encode\_SiP.to.RPG}, we are interested in designing an efficient and
easily implementable algorithm for decoding
the permutation~$\pi^*$ from the graph~$F[\pi^*]$. Thus, we next present such a decoding algorithm,
we call it {\tt Decode\_RPG.to.SiP}, which
is easily implementable: indeed, the only operations used by
the algorithm are edge modifications on $F[\pi^*]$
and DFS-search on trees.

The algorithm takes as input a reducible permutation
flow-graph~$F[\pi^*]$ on $n^*+2$ nodes constructed by
Algorithm {\tt Encode\_SiP.to.RPG}, and produces
a permutation~$\pi^*$ of length $n^*$;
it works as follows:


\vspace*{0.15in} \noindent Algorithm {\tt Decode\_RPG.to.SiP}
\vspace*{-0.04in}
\begin{enumerate}
\item[1.\,]   
Delete the forward edges $(u_{i}, u_{i-1})$ from
the set $E(F[\pi^*])$, $1 \leq i \leq n^*+1=s$,
and the node $t = u_0$ from $V(F[\pi^*]) = \{s = u_{n^*+1}, u_{n^*}, \ldots, u_1, u_0 = t\}$;
%
\item[2.\,]   
Flip all the remaining edges (backward edges) of the
graph~$F[\pi^*]$ yielding the Dmax-tree $T_{d}[\pi^*]$
with nodes $s, u_1, u_2, \ldots, u_{n^*}$;
%
\item[3.\,]   
Perform DFS-search on the tree $T_{d}[\pi^*]$ starting at node~$s$
by always proceeding to the minimum-labeled child node
and compute the DFS discovery time $d[u]$ of each node~$u$
of $T_{d}[\pi^*]$;
%
\item[4.\,]   
Order the nodes $u_1, u_2, \ldots, u_{n^*}$ of the tree~$T_{d}[\pi^*]$
by their DFS discovery time $d[]$ and
let $\pi=(u'_1, u'_2, \ldots, u'_{n^*})$ be the resulting order;
%
\item[5.\,]   
Return $\pi^* = \pi$;
\end{enumerate}

\vspace*{0.04in}
\noindent
{\it Time and Space Complexity.}
The size of the reducible permutation graph~$F[\pi^*]$ constructed
by the algorithm {\tt Encode\_SiP.to.RPG} is $O(n^*)$, where
$n^*$ is the length of the permutation $\pi^*$, and
thus the size of the resulting tree~$T_{d}[\pi^*]$ is also $O(n^*)$.
It is well known that the DFS-search on the tree~$T_{d}[\pi^*]$
takes time linear in the size of $T_{d}[\pi^*]$.
Thus, the decoding algorithm is executed in $O(n^*)$ time
using $O(n^*)$ space. Thus, the following theorem holds:

\vspace*{0.04in}
\begin{theorem}\label{thm:theo5.2}
Let $F[\pi^*]$ be a reducible permutation flow-graph of size $O(n^*)$
produced by the algorithm {\tt Encode\_SiP.to.RPG}.
The algorithm {\tt Decode\_RPG.to.SiP} decodes
the flow-graph~$F[\pi^*]$ in $O(n^*)$ time and space.
\end{theorem}

\vspace*{0.1in}
\section{Structure and Properties of the SiP~$\pi^*$}
\label{sec:Properties-SiP}
In this section, we analyze the structure of a self-inverting
permutation~$\pi^*$ produced by the algorithms {\tt Encode\_W.to.SiP}
and present properties which are important in their own right,
at least from a graph-theoretic point of view, and prove useful
in shielding our watermark graph~$F[\pi^*]$ against attacks.

\vspace*{0.1in}
\subsection{The Subsequences $\pi_1^*$ and $\pi_2^*$}
\label{subsec:P1P2}

Consider a self-inverting permutation~$\pi^*$ encoding an integer $w$ in the range $[2^{n-1},2^n-1]$, where $n$ is the length of the binary representation of $w$; we distinguish the following two cases:

\medskip\noindent
{\bf All-One case}: Suppose that $w = 2^n-1$, that is, all the bits in
$w$'s binary representation are $1$.
Then, according to Algorithm Encode\_W.to.SiP,
$B' = 0^n 1^n 0$,
$\pi^b = (n+1, \, n+2, \, \ldots, \, 2n+1, \, n,
\, n-1, \, \ldots, \, 2, \, 1)$, and
$\pi^* = (n+1, \, n+2, \, \ldots, \, 2n,
\, 1, \, 2, \, \ldots, \, n, \, 2n+1)$, that is, $\pi^*$ is the concatenation of
the (increasing) subsequences $\pi^*_1$ and $\pi^*_2$
where
\begin{equation} \label{eq:eq1}
\pi^*_1 \ =\  (n+1, \, n+2, \, \ldots, \, 2n)
\qquad \hbox{and} \qquad
\pi^*_2 \ =\  (1, \, 2, \, \ldots, \, n, \, 2n+1).
\end{equation}

\medskip\noindent
{\bf Zero-and-One case}: Suppose that $w \neq 2^n-1$. Then, the concatenation of
the binary representation of $w$ with a trailing~$0$ consists of
$a_1$ $1$s, followed by $b_1$ $0$s, followed by $a_2$ $1$s,
followed by $b_2$ $0$s, and so on, followed by $a_\ell$ $1$s,
followed by $b_\ell$ $0$s where $\ell \ge 1$, $a_i, b_i > 0$,
and $a_1 < n$.

For convenience, let
$A_j = \sum_{t=1}^{j} a_t$ and
$\Gamma_j = \sum_{t=1}^{j} (a_t+b_t)$ for $0 \le j \le \ell$;
note that $A_0 =0$, $\Gamma_0 = 0$, $\Gamma_\ell = 2n+1$,
and $A_\ell$ is equal to the number of $1$s in
the binary representation of $w$.
Additionally, let $B_j = \sum_{t=1}^{j} b_t$ for $0 \le j \le \ell$,
and $\overline{B}_j = \left( \sum_{t=1}^{\ell} b_t \right) - B_j$,
which imply that $B_0 = \overline{B}_\ell = 0$,
$\overline{B}_{j-1} = \overline{B}_{j} + b_j$,
and $\Gamma_\ell = A_\ell + B_\ell = A_\ell + \overline{B}_0$.

Then, according to Algorithm Encode\_W.to.SiP,
$B' \, = \, 0 \cdots 0 || B || 0$ where $B$ is
the binary representation of $w$, and
$\pi^b \ = \ X \, || \, Y^R$
where $X$ is the sequence of the $1$s' positions in $B'$
and $Y^R$ is the reverse of the sequence of the $0$s' positions
in $B'$, that is,
\begin{eqnarray*}
X & = & X_1 \, || \, X_2 \, || \, \ldots \, || \, X_i
 \, || \, \ldots \, || \, X_{\ell}\cr
& = & \Bigl(
\underbrace{n+1, \ldots, n+a_1}_{a_1},
\  \underbrace{n+\Gamma_1+1, \ldots, n+\Gamma_1+a_2}_{a_2},
\  \ldots,
\  \underbrace{n+\Gamma_{i-1}+1, \ldots,
n+\Gamma_{i-1}+a_i}_{a_i},\cr
& \ & \phantom{x} \ldots, \  \underbrace{n+\Gamma_{\ell-1}+1, \ldots,
n+\Gamma_{\ell-1}+a_\ell}_{a_\ell}
\Bigr)
\end{eqnarray*}
and
\begin{eqnarray*}
Y^R & = & Y_\ell^R \, || \, \ldots \, || \, Y_i^R \, ||
\, \ldots \, || \, Y_1^R \, || \, Y_0^R\cr
& = & \Bigl(
\underbrace{n+\Gamma_\ell, \ldots, n+\Gamma_\ell-b_\ell+1}_{b_\ell},
\  \ldots,
\  \underbrace{n+\Gamma_i, \ldots, n+\Gamma_i-b_i+1}_{b_i},
\  \ldots,\cr
& \  & \phantom{x}
\underbrace{n+\Gamma_1, \ldots, n+\Gamma_1-b_1+1}_{b_1},
\  \underbrace{n, \, n-1, \ldots, 2, \, 1}_{n}
\Bigr).
\end{eqnarray*}
(Note that $\Gamma_{i-1}+a_i = \Gamma_i-b_i$.)

Next, because the total length of the concatenation of
$X_1, X_2, \ldots, X_\ell, Y_\ell^R, \ldots, Y_1^R$ is $n+1$
whereas the length of $Y_0^R$ is $n$,
the cycles constructed in Step~5 of Algorithm Encode\_W.to.SiP
are:
\begin{itemize}
\item
\textit{2-cycles}: the 2-cycles in increasing order of their second elements from $1$ to $n$ are (note that $A_0 = \overline{B}_\ell = \Gamma_0 = 0$,
$A_\ell+\overline{B}_0-1 = \Gamma_\ell-1 = (n+1) - 1 = n$,
and $\Gamma_{i-1}+a_i = \Gamma_i - b_i$):\\
\phantom{x}
$(n+1, \ A_0+1)$, \ $(n+2, \ A_0+2)$, \ $\ldots$,
\ $(n+a_1, \ A_1)$,\\
\phantom{x}
$(n+\Gamma_1+1, \ A_1+1)$, \ $(n+\Gamma_1+2, \ A_1+2)$, \ $\ldots$,
\ $(n+\Gamma_1+a_2, \ A_2)$,\\
\phantom{x}
$\cdots$\\
\phantom{x}
$(n+\Gamma_{i-1}+1, \ A_{i-1}+1)$,
\ $(n+\Gamma_{i-1}+2, \ A_{i-1}+2)$, \ $\ldots$,
\ $(n+\Gamma_{i-1}+a_i, \ A_i)$,\\
\phantom{x}
$\cdots$\\
\phantom{x}
$(n+\Gamma_{\ell-1}+1, \ A_{\ell-1}+1)$,
\ $(n+\Gamma_{\ell-1}+2, \ A_{\ell-1}+2)$, \ $\ldots$,
\ $(n+\Gamma_{\ell-1}+a_\ell, \ A_\ell)$,\\
\phantom{x}
$(n+\Gamma_\ell, \ A_\ell+ \overline{B}_{\ell}+1)$,
\ $(n+\Gamma_\ell-1, \ A_\ell+ \overline{B}_{\ell}+2)$, \ $\ldots$,
\ $(n+\Gamma_\ell-b_\ell+1, \ A_\ell+ \overline{B}_{\ell-1})$,\\
\phantom{x}
$\cdots$\\
\phantom{x}
$(n+\Gamma_i, \ A_\ell+ \overline{B}_{i}+1)$,
\ $(n+\Gamma_i-1, \ A_\ell+ \overline{B}_{i}+2)$, \ $\ldots$,
\ $(n+\Gamma_i-b_i+1, \ A_\ell+ \overline{B}_{i-1})$,\\
\phantom{x}
$\cdots$\\
\phantom{x}
$(n+\Gamma_2, \ A_\ell+ \overline{B}_{2}+1)$,
\ $(n+\Gamma_2-1, \ A_\ell+ \overline{B}_{2}+2)$, \ $\ldots$,
\ $(n+\Gamma_2-b_2+1, \ A_\ell+ \overline{B}_1)$,\\
\phantom{x}
$(n+\Gamma_1, \ A_\ell+ \overline{B}_{1}+1)$,
\ $(n+\Gamma_1-1, \ A_\ell+ \overline{B}_{1}+2)$, \ $\ldots$,
\ $(n+\Gamma_1-b_1+2, \ A_\ell+ \overline{B}_0-1)$;
\item
\textit{1-cycle}: the 1-cycle involves the last element of $Y_1^R$,
that is, it is $(n + \Gamma_1 - b_1 + 1) = (n + a_1 + 1)$.
\end{itemize}

\noindent
Therefore, the self-inverting permutation $\pi^*$ is
the concatenation of $\pi^*_1$ and $\pi^*_2$, where
\begin{equation}\label{eq:eq2}
\pi^*_1 \  = \  X_1 \, || \, X_2 \, || \, \ldots \, ||
\, X_\ell \, || \, Y_\ell^R \, || \, \ldots \, || \, Y_2^R
\, || \, \Psi^R_1
\end{equation}
with $\Psi^R_1 = (n+\Gamma_1, \ldots, n+\Gamma_1-b_1+2)$
(note that $\Psi^R_1$ is empty if $b_1 = 1$ otherwise
it is equal to $Y_1^R$ without its last element
$n+\Gamma_1-b_1+1 = n + a_1 + 1$)
and
\begin{eqnarray}\label{eq:eq3}
\pi^*_2 & = & \Bigl( \underbrace{1, \, 2, \, \ldots, \, A_1}_{a_1},
\ n+A_1+1,
\ \underbrace{n, n-1, \ldots, n-B_1+2}_{b_1-1},\cr
& \ & \phantom{x}
\underbrace{A_1+1, \, A_1+2, \, \ldots, \, A_2}_{a_2},
\ \underbrace{n-B_1+1, \, n-B_1, \ldots, n-B_2+2}_{b_2},\cr
& \ & \phantom{x} \cdots,\cr
& \ & \phantom{x}
\underbrace{A_{i-1}+1, \, A_{i-1}+2, \, \ldots, \, A_i}_{a_i},
\ \underbrace{n-B_{i-1}+1, \, n-B_{i-1}, \ldots, n-B_i+2}_{b_i},
\cr
& \ & \phantom{x} \cdots,\cr
& \ & \phantom{x}
\underbrace{A_{\ell-1}+1, \, A_{\ell-1}+2, \, \ldots,
\, A_\ell}_{a_\ell},
\ \underbrace{n-B_{\ell-1}+1, \, n-B_{\ell-1}, \ldots,
 n-B_\ell+2}_{b_\ell} \Bigr)
\end{eqnarray}
(note that the last element of $\pi^*_2$ is
$n-B_\ell+2 = (\Gamma_\ell-1) - B_\ell + 2 =
(A_\ell + B_\ell-1) - B_\ell + 2 = A_\ell+1$).
It is interesting to note that $\pi^*_1$ is a permutation of
the numbers $n+1, n+2, \ldots, 2 n+1$ except for
$n+a_1+1$; in turn, $\pi^*_2$ is a permutation of
the numbers $1, 2, \ldots, n$ and $n+a_1+1$.
Additionally, $\pi^*_2$ consists of the $a_1$ numbers
$1,2,\ldots,A_1$ followed by $b_1$ numbers larger than $A_\ell+1$,
followed by the $a_2$ numbers $A_1+1, A_1+2, \ldots, A_2$,
followed by $b_2$ numbers larger than $A_\ell+1$,
and so on, up to the $a_\ell$ numbers
$A_{\ell-1}+1, A_{\ell-1}+2, \ldots, A_\ell$ that
are followed by the $b_\ell$ numbers
$A_\ell+b_\ell, A_\ell+b_\ell-1, \ldots, A_\ell$
(note that $n-B_{\ell-1}+1 = (\Gamma_\ell-1) - B_{\ell-1} + 1 =
(A_\ell + B_\ell-1) - B_{\ell-1} + 1 =
A_\ell + (B_\ell - B_{\ell-1}) = A_\ell + b_\ell$).

\bigskip\noindent
\textbf{Example 4.1}
Consider $w = 220$.
The binary representation of $w$ is $11011100$,
and hence $n = 8$.
From the concatenation of the binary representation of $w$
with the additional trailing $0$, we have that
$\ell = 2$, $a_1 = 2$, $b_1 = 1$, $a_2 = 3$, $b_2 = 3$;
in turn, $A_1 = a_1 = 2$, $A_2 = a_1 + a_2 = 5$,
$B_1 = \overline{B}_1 = b_1 = 1$,
$B_2 = \overline{B}_0 = b_1 + b_2 = 4$,
$\Gamma_1 = A_1 + B_1 = 3$, and $\Gamma_2 = A_2 + B_2 = 9$.
Then,
$$\pi^b =
(\underbrace{0, \, 0, \, 0, \, 0, \, 0, \, 0, \, 0, \, 0}_{n=8},
\, \underbrace{1, \, 1, \, 0, \, 1, \, 1, \, 1, \, 0, \, 0}_{n=8},
\, 0)$$
and
$\pi^*_1 = (9, \, 10, \, 12, \, 13, \, 14, \, 17, \, 16, \, 15)$
and
$\pi^*_2 = (1, \, 2, \, 11, \, 3, \, 4, \, 5, \, 8, \, 7, \, 6)$;
thus,
$$\pi^* = (9, \, 10, \, 12, \, 13, \, 14, \, 17, \, 16, \, 15,
\, 1, \, 2, \, 11, \, 3, \, 4, \, 5, \, 8, \, 7, \, 6).$$

\vspace*{0.1in}
\subsection{The 4-Chain Property}
\label{subsec:4chain}

Based on the structure of a self-inverting permutation~$\pi^*$
produced by algorithm {\tt Encode\_W.to.SiP}, we next present four important
properties of $\pi^*$ which are incorporated into our codec watermark graph $F[\pi^*]$ making it resilient against attacks.

\begin{enumerate}
\item[$\bullet$\,]
{\bf Odd-One property}: The self-inverting permutation $\pi^*$ produced by the encoding algorithm
{\tt Encode\_W.to.SiP} has always {\it odd length} and contains {\it exactly one cycle} of length 1.

\vspace*{0.04in}
\item[$\bullet$\,]
{\bf Bitonic property}:
The self-inverting permutation~$\pi^*$ is constructed from the bitonic sequence $\pi^b = X || Y^R$, where $X$ and $Y$
are increasing subsequences (see Step~4 of our encoding algorithm {\tt Encode\_W.to.SiP}), and thus the bitonic property of $\pi^b$
is encapsulated in $\pi_1^*$. Indeed, the first $n$ elements of permutation~$\pi^*$
form the bitonic sequence $\pi_1^*$ ($\pi_1^*$ first monotonically increases and then monotonically decreases or simply monotonically increases).
We say that the SiP $\pi^*$ has the Bitonic property if $\pi_1^*$ is a bitonic sequence.

It is easy to see that there exists SiPs which have the Odd-One property but do not satisfy the Bitonic property; consider, for example, the permutation $\pi=(2,1,4,3,6,5,7)$.

\vspace*{0.04in}
\item[$\bullet$\,]
{\bf Block property}: The algorithm {\tt Encode\_W.to.SiP}
takes the binary representation of the integer~$w$ and
initially constructs the binary number $B'$ (see Step~2).
The binary representation of $B' = \underbrace{00 \cdots 0}_n||B||0$ consists of three parts
(or blocks):
    \begin{enumerate}
    \item[(i)] the first part contains the leftmost $n$ bits,
each equal to $0$,
    \item[(ii)] the second part contains the next $n$ bits which
form the binary representation $B$ of the integer~$w$, and
    \item[(iii)] the third part of length 1 contains a bit $0$.
    \end{enumerate}
The structure of $B'$ affects the construction of both subsequences $X$
and $Y$ (see Bitonic property), and thus the elements of $\pi_1^*$, i.e.,
the first $n$ elements of permutation~$\pi^*$, have values in the set $H=\{n+1, n+2, \ldots, 2n, 2n+1\}$.
Consequently, we say that the SiP $\pi^*$ has the Block property if all the elements of $\pi_1^*$ belong to $H$.

Since $|\pi_1^*|=n$ and $|H|=n+1$,
there is one element $\alpha \in H$ which does not participate
in $\pi_1^*$. Moreover, since $\pi^*$ is a SiP and
its first $n$ elements have values greater that $n$,
it follows that the element~$\alpha$ forms
the 1-cycle $(\alpha, \alpha)$ of $\pi^*$.
For example, consider the SiP $\pi^* = (5, 6, 9, 8, 1, 2, 7, 4, 3)$
which encodes the integer $w=12$ with binary representation $B = 1100$;
the element~$\alpha$ has value $7$ since $n=4$, $H=\{5, 6, 7, 8, 9\}$,
and $\pi_1^*=(5, 6, 9, 8)$.

Notice that there exist SiPs having the Bitonic property which do not satisfy the Block property; for example, $\pi=(5,6,9,4,1,2,8,7,3)$ is such a SiP.

\vspace*{0.04in}
\item[$\bullet$\,]
{\bf Range property}: Let $R_n$ denote the range of all the integers $w$ having $n$-bit representation
with the most significant bit (msb)
equal to 1. Thus, $R_n=[2^{n-1}, 2^n-1]$, $n \geq 1$.

For any integer $w \in R_n$, the $n$-bit representation of $w$ has msb=1 and thus the first element of
sequence $X$ is equal to $n+1$ (see Step 3 of algorithm {\tt Encode\_W.to.SiP}).
Then, by construction, the first element of the SiP $\pi^*$ has value $n+1$. We say that a SiP $\pi^*$
has the Range property if the first element of $\pi_1^*$ has value $n+1$; in this case, by definition, the first element
of $\pi_2^*$ is equal to $1$.

Note: Clearly we can use $n$ bits to represent an integer
$w' \in R_{n'}$, where $n > n'$ (for example,
the $4$-bit representation of $w'=5 \in R_3$ is the binary number $B=0101$); then, the msb is equal to 0.
In this case, our algorithm {\tt Encode\_W.to.SiP} works correctly and produces a SiP $\pi^*$ of length $n^*=2n+1$
which has both the Bitonic property and the Block property, but does not have the Range property.
For example, for $w'=5$ and $n=4$ our algorithm produces the SiP $\pi^*=(6,8,9,7,5,1,4,2,3)$.

\end{enumerate}

\vspace*{0.05in}
\noindent
\textbf{Observation 4.1}
Let $w$ be an integer in $R_n=[2^{n-1}, 2^n-1]$ and
let $\pi^*$ be the SiP of length~$n^*=2n+1$ produced by
our encoding algorithm {\tt Encode\_W.to.SiP}.
By construction, in all the cases the maximum element~$2n+1$ of $\pi^*$
participates in $\pi_1^*$, except for the case where $w$ is
the last integer in the range $R_n$, i.e., $w=2^n-1$.
In that case, since the binary representation of $w$ is
$B=11 \cdots 1$, the bitonic sequence~$\pi_1^*$ is
$(n+1, n+2, \ldots, 2n)$, which is trivially bitonic
since it monotonically increases.
It follows that $\alpha$ is the maximum element~$2n+1$, which
is located in the last position of $\pi^*$;
see the All-One case analysis in Subsection~\ref{subsec:P1P2}.
We can see the structure of $\pi^*$ by considering, for example,
the encoding of integer $w=15 \in R_4$, where $B=1111$ and
thus $\pi^*=(5,6,7,8,1,2,3,4,9)$.

\vspace*{0.05in}
\noindent
Hereafter, the term SiP will refer to a self-inverting permutation over the set $N_{n^*}$ having
the above four properties.

\vspace*{0.1in}
\subsection{The Structure of the SiP}
\label{subsec:CiS-Property}

Let $w \in R_n=[2^{n-1}, 2^n-1]$ be an integer encoded by
the SiP~$\pi^*=\pi_1^*||\pi_2^*$ where $\pi_1^*$ and $\pi_2^*$
are subsequences of lengths $n$ and $n+1$, respectively;
see Subsection~\ref{subsec:P1P2}.
As in Subsection \ref{subsec:4chain}, $\alpha$ denotes the element
of $\pi^*$ which forms its 1-cycle.
Moreover, by $\beta$ and $\gamma$ we denote the last elements
of $\pi_1^*$ and $\pi_2^*$, respectively.

Based on the structures of the two subsequences $\pi_1^*$ and
$\pi_2^*$, we conclude that the structure of the SiP
$\pi^*=\pi_1^*||\pi_2^*$ has the following two forms:
    \begin{enumerate}
    \item[($i$)] {\it All-One case}: $w = 2^n-1$.
In this case, the SiP $\pi^*$ is the concatenation of
two increasing sequences:
    $$\pi_1^*=(n+1, n+2, \ldots, 2n) \ \ \ \text{and} \ \ \ \pi_2^*=(1, 2, \ldots, n, 2n+1)$$
    where the maximum element $2n+1$ of $\pi^*$ belongs to $\pi_2^*$ and,
    since $\pi^*=\pi_1^* || \pi_2^*$, it is located in the last position
    of $\pi^*$, i.e., $\alpha=\gamma=2n+1$.

    \item[($ii$)] {\it Zero-and-One case}: $w \in \widehat{R}_n=[2^{n-1}, 2^n-2]$.
    In this general case, the structure of $\pi^*$ is of the form:
    $$\pi_1^*=(n+1, n+2, \ldots, n+k, n+k_1, \ldots, 2n+1, \ldots, \beta)
    \ \ \text{and} \ \ \pi_2^*=(1, 2, \ldots, k, \alpha, \ldots, \gamma)$$
    where $k \geq 1$ and $k_1 > k+1$. Moreover, $\alpha = n+k+1$.
    Notice that the All-One case follows from the general Zero-and-One case by setting $k=n$.

    \end{enumerate}

\noindent
Recall that $\pi_1^*$ and $\pi_2^*$ are of lengths $n$ and $n+1$,
respectively, that the sequence~$\pi_1^*$ is bitonic whose elements
have values greater than $n$ and its first element always has value
$n+1$, and that the sequence~$\pi_2^*$ always contains
the element~$\alpha$ and its first element has value 1.

\vspace*{0.1in}
\noindent {\bf Increasing Subsequences}:
Let $\pi^*=(\pi_{1}, \pi_{2}, \ldots, \pi_{n^*})$ be a SiP of
length~$n^* = 2n+1$ produced by our encoding algorithm
{\tt Encode\_W.to.SiP} and let $I_1$, $I_2$, $\ldots$, $I_h$ be
the $1$st, $2$nd, $\ldots$, $h$th increasing subsequence of $\pi^*$,
respectively; see Subsection \ref{subsec:Preliminaries}.

(i) In the All-One case, where the number $w = 2^n-1$ is encoded,
the SiP~$\pi^*$ has two increasing subsequences $I_1$ and $I_2$
having the following form:
$$I_1 =(n+1, n+2, \ldots, 2n, 2n+1) \ \ \ \text{and}
\ \ \ \ I_2 =(1, 2, \ldots, n)$$
that is, $I_1 = \pi_1^* || (2n+1)$ and
$I_2 = \pi_2^* \backslash (2n+1)$.

(ii) Let us now consider the Zero-and-One case, where
$w \in \widehat{R}_n=[2^{n-1}, 2^n-2]$, and
let $\pi_1^*=\pi_{11}^*||\pi_{12}^*$ and
$\pi_2^*=\pi_{21}^*||\pi_{22}^*$ where:
    $$\pi_{11}^*=(n+1, \ldots, 2n+1), \ \pi_{12}^*=(p_1, p_2, \ldots, \beta)
\ \ \text{and} \ \ \pi_{21}^*=(1, 2, \ldots, k, \alpha), \ \pi_{22}^*=(q_1, q_2, \ldots, \gamma).$$

From the structure of subsequences $\pi_1^*$ and $\pi_2^*$
and since $\pi^*$ is a SiP,
it follows that $\pi_{11}^*$ is an increasing sequence,
$\pi_{12}^*$ is a decreasing sequence (because $\pi_1^*$ is bitonic)
whose minimum element~$\beta$ is larger than all the elements
in $\pi_{21}^*||\pi_{22}^*$,
$\pi_{21}^*$ is an increasing sequence whose maximum element~$\alpha$
is larger than all the elements in $\pi_{22}^*$, and
$\gamma$ is equal to the index of the maximum element~$2n+1$
in $\pi^*$; see Equation~\ref{eq:eq3}.

We next focus on the structure of the sequence
$\pi_{22}^*=(q_1, q_2, \ldots, \gamma)$.
To this end, we rewrite the sequence $\pi_1^*$ more analytically
as the concatenation of three subsequences,
that is, $\pi_1^*=\pi_{1a}^*||\pi_{1b}^*||\pi_{1c}^*$ such that
$$\pi_{1a}^*= (n+1, n+2, \ldots, n+k), \ \ \pi_{1b}^* = (n+k_1, n+k_2, \ldots, n+k_m), \ \ \pi_{1c}^* = (2n+1, p_1, p_2, \ldots, \beta)$$
where $k \geq 1$ and $m \geq 0$.
By construction:

\medskip\medskip
\noindent {\sc Fact 4.1.} {\it The indices of the elements of the subsequence
$\pi_{1a}^* = (n+1, n+2, \ldots, n+k)$
form the sequence $\pi_{21}^* \backslash (\alpha) = (1, 2, \ldots, k)$.}

\smallskip\medskip\noindent
Indeed, $n+1$ is the 1st element of $\pi^*$, $n+2$ is the 2nd element,
while $n+k$ is the $k$th element of $\pi^*$ and the length of
$\pi_1^*$ is $n$.
Additionally,

\medskip\medskip
\noindent {\sc Fact 4.2.} {\it The indices of the elements of the subsequence
$$\pi_{1b}^*||\pi_{1c}^*=(n+k_1, n+k_2, \ldots, n+k_m, 2n+1, p_1, p_2, \ldots, \beta)$$
form the sequence $\pi_{22}^*=(q_1, q_2, \ldots, \gamma)$, where $\gamma$ is the index of the maximum element $2n+1$.}

\smallskip\medskip\noindent
The indices of the elements of $\pi_{22}^*$ are in the range $[k+1, n]$;
indeed, $n+k_1$ is the $(k+1)$st element of $\pi^*$, $n+k_2$ is the $(k+2)$nd element,
while its last element $\beta$ is the $n$th element of $\pi^*$. Thus, the indices of the subsequence $\pi_{1b}^*||\pi_{1c}^*$ have the following form:
$$(k+1, k+2, \ldots, k+m) \ \ || \ \ (\gamma, \gamma+1, \gamma+2, \ldots, n)$$
Since the sequence of indices
$(\gamma, \gamma+1, \gamma+2, \ldots, n)$ corresponds to
the elements of the decreasing sequence
$\pi_{1c}^*=(2n+1, p_1, p_2, \ldots, \beta)$, it follows that
these indices will appear in $\pi_{22}^*$ in reverse order,
i.e., $n, n-1, \ldots, \gamma$.
Moreover, since the sequence of indices
$(k+1, k+2, \ldots, k+m)$ corresponds to the elements of
the increasing sequence $\pi_{1b}^*=(n+k_1, n+k_2, \ldots, n+k_m)$,
these indices will appear in $\pi_{22}^*$ in increasing order,
i.e., $k+1, k+2, \ldots, k+m$. Moreover, the maximum element~$k+m$
in $(k+1, k+2, \ldots, k+m)$ is less than the min element~$\gamma$
in $(\gamma, \gamma+1, \gamma+2, \ldots, n)$.

Let $\lambda$ be the length of the sequence $\pi_{1c}^*=(2n+1, p_1, p_2, \ldots, \beta)$.
Then the  sequence $\pi_{22}^*$ is the concatenation of
$\lambda$ increasing subsequences $\Phi_1, \Phi_2, \ldots, \Phi_{\lambda}$
with last elements $n, n-1, \ldots, \gamma$, respectively.
That is, $\pi_{22}^*$ has the following form:
$$\pi_{22}^* \ =
\ \Phi_1 \ || \ \Phi_2 \ || \ \cdots \ || \ \Phi_{\lambda} \ \ =
\ \ (\sigma_1, n) \ || \ (\sigma_2, n-1) \ || \ \cdots \ ||
\ (\sigma_{\lambda}, \gamma)$$
where $\sigma_1, \sigma_2, \ldots, \sigma_{\lambda}$ are subsequences
(some of which may be empty) such that
$\sigma_1 \,||\, \sigma_2 \,||\, \ldots \,||\, \sigma_{\lambda}
=(k+1, k+2, \ldots, k+m)$.

Having analyzed the structure of the four subsequences
$\pi_{11}^*$, $\pi_{12}^*$, $\pi_{21}^*$, and $\pi_{22}^*$
in the case where $w \in \widehat{R}_n=[2^{n-1}, 2^n-2]$, along with
the subsequences $\pi_{1a}^*$, $\pi_{1b}^*$ and $\pi_{1c}^*$,
we can then easily get the structure of the increasing subsequences
of $\pi^*$. It is easy to see that $\pi^*$ is the concatenation of
$2\lambda+1$ consecutive increasing subsequences:
$$\pi^* \ = \ I_1 \ ||  \ I_2 \ || \ \ldots \ || \ I_{\lambda} \ ||
\ I_c \ || \ I_{\lambda+1} \ || \ I_{\lambda+2} \ || \ \ldots \ ||
\ I_{2\lambda}$$
where
$I_1=(n+1, n+2, \ldots, 2n+1)$,
$I_2=(p_1)$, $I_3=(p_2)$, $\ldots$, $I_\lambda=(\beta)$,
$I_c=(1, 2, \ldots, k, \alpha)$,
$I_{\lambda+1}=\Phi_1$, $I_{\lambda+2}=\Phi_2$, $\ldots$,
$I_{2\lambda}=\Phi_\lambda$. Hereafter, the sequence
$$I(\pi^*)=[I_1, \ I_2, \ \ldots, \ I_{\lambda}, \ I_c, \ I_{\lambda+1}, \ \ldots, \ I_{2\lambda}]$$
will be referred to as the {\it increasing-subsequence representation}
or, for short, {\it $I$-representation} of $\pi^*$.

Additionally, Facts 4.1 and 4.2 and since the first element of $\pi_2^*$ is equal to 1 imply the following lemma.

\vspace*{0.04in}
\begin{lemma}\label{lemma:SiP_fix}
Let $\pi^*$ be a self-inverting permutation of length~$n^* = 2n+1$ produced by algorithm {\tt Encode\_W.to.SiP} and let $\pi'$ be the sequence resulting from $\pi^*$ after either its leftmost $n$ elements or its rightmost $n$ elements have been deleted.
The SiP~$\pi^*$ can be fully reconstructed from $\pi'$.
\end{lemma}


\bigskip\noindent
\textbf{Example 4.2}
\ Consider the SiP $\pi^* = (8, \, 9, \, 10, \, 11, \, 12, \, 13,
\, 14, \, 1, \, 2, \, 10, \, 3, \, 7, \, 6, \, 4, \, 15)$ which
encodes the number $w = 127$, that is, the last number in
the range $R_7 = [2^{n-1}, 2^n-1] = [64, 127]$; the binary
representation of $w$ is $1111111$ and thus $n=7$ (case All-One).
In this case, the two subsequences $\pi_{1}^*$ and $\pi_{2}^*$ are:
$$\pi_{1}^*=(8, 9, 10, 11, 12, 13, 14) \ \ \ \text{and}
\ \ \ \pi_{2}^*=(1, 2, 3, 4, 5, 6, 7, 15)$$
where $\alpha=\gamma=2n+1=15$. The SiP~$\pi^*$ consists of
two increasing subsequences $I_1 = \pi_1^* || (2n+1)$ and
$I_2 = \pi_2^* \backslash (2n+1)$ and, thus,
its $I$-representation $I(\pi^*) = [I_1, I_2]$ is:
$$I(\pi^*) = [(8, 9, 10, 11, 12, 13, 14,15),
\ (1, 2, 3, 4, 5, 6, 7)].$$

\bigskip\noindent
\textbf{Example 4.3}
\ Consider now the SiP $\pi^* = (8, \, 9, \, 11, \, 14, \, 15, \, 13,
\, 12, \, 1, \, 2, \, 10, \, 3, \, 7, \, 6, \, 4, \, 5)$ which
encodes the number $w = 105$; the binary representation of $w$ is
$1101001$ and $w$ belongs to the range
$\widehat{R}_7 = [2^{n-1}, 2^n-2] = [64, 126]$ (case Zero-and-One).
The subsequences $\pi_{11}^*$, $\pi_{12}^*$, $\pi_{21}^*$, and
$\pi_{22}^*$ are:
$$\pi_{11}^*=(8, 9, 11, 14, 15), \ \ \pi_{12}^*=(13, 12),
\ \ \pi_{21}^*=(1, 2, 10) \ \ \text{and}
\ \ \pi_{22}^*=(3, 7, 6, 4, 5)$$
where $\alpha=10$, $\beta=12$, $\gamma=5$, and the maximum element of
$\pi^*$ is $2n+1=15$. Since the decreasing subsequence $(15, 13, 12)$
has length $\lambda=3$, it follows that their indices $5, 6, 7$
appear in $\pi_{22}^*$ in reverse order and thus $\pi_{22}^*$ is
the concatenation of 3 increasing subsequences $\Phi_1, \Phi_2$,
and $\Phi_3$ with last elements $7, 6, 5$, respectively;
indeed, $\Phi_1=(3,7)$, $\Phi_2=(6)$ and $\Phi_3=(4,5)$.
Thus, the whole SiP~$\pi^*$ is the concatenation of $2\lambda+1=7$
increasing subsequences
$$\pi^* = I_1 \ || \ I_2 \ || \ I_3 \ || \ I_c \ || \ I_4 \ ||
\ I_5 \ || \ I_6$$
and has the following $I$-representation:
$$I(\pi^*) = [(8,9,11,14,15), \ (13), \ (12), \ (1,2,10), \ (3,7),
\ (6), \ (4,5)].$$

\bigskip\noindent
\textbf{CiS and 2iS Properties:}
\ Let $\pi$ be a permutation over the set $N_{2n+1}$, $n \geq 1$,
and let $S_1$, $S_2$, $\ldots$, $S_k$ ($k \ge 1$) be
the $1$st, $2$nd, $\ldots$, $k$th increasing subsequences of $\pi$,
respectively; see Subsection~\ref{subsec:Preliminaries}.
We say that the permutation~$\pi$ has
the {\it consecutive increasing subsequence property}
(or CiS property, for short)
if $\pi= S_1 || S_2 || \cdots || S_k$, $k \geq 1$.
Additionally, we say that $\pi$ has
the {\it two increasing subsequence property}
(or 2iS property, for short) if it has 2 increasing subsequences
$S_1= (n+1, n+2, \ldots, 2n, 2n+1)$ and
$S_2= (1, 2, \ldots, n)$ and
$\pi= \Bigl( S_1 \backslash (2n+1) \Bigr) \ || \ S_2 \  || \  (2n+1)$.
For example, the permutation
$\pi^* = (5, \, 6, \, 9, \, 8, \, 1, \, 2, \, 7, \, 4, \, 3)$
satisfies the CiS property since $S_1=(5, \, 6, \, 9)$,  $S_2=(8)$,
$S_3=(1, \, 2, \, 7)$, $S_4=(4)$ and $S_5=(3)$, and
$\pi= S_1 || S_2 || \ldots || S_5$, whereas
the permutation $\pi^* = (5, \, 6, \, 7, \, 8, \, 1, \, 2, \, 3,
\, 4, \, 9)$ satisfies the 2iS property.

\medskip
Our analysis of the structure of the permutation~$\pi^*$ encoding
a number~$w$ in the range $R_n=[2^{n-1}, 2^n-1]$ with respect to its increasing subsequences establishes that $\pi^*$ has the 2iS property
in the case where $w = 2^n-1$ (case All-One) and the CiS property
in the case where $w \in \widehat{R}_n=[2^{n-1}, 2^n-2]$
(case Zero-and-One); see also Examples 4.2 and 4.3.
Thus, we can state the following result.

\vspace*{0.04in}
\begin{theorem}\label{thm:2iS-CiS-properties}
Let $\pi^*$ be the SiP which encodes a number $w$ of the range $R_n=[2^{n-1}, 2^n-1]$ and let $I_1, \ I_2, \ldots, I_{k}$ be the $1$st, $2$nd, $\ldots$, $k$th increasing subsequences of $\pi^*$, respectively. Then:
    \begin{enumerate}
    \item[($i$)] if $w \in \widehat{R}_n=[2^{n-1}, 2^n-2]$,
$\pi^*$ satisfies the CiS property, i.e.,
$\pi^*= I_1 \,|| \,I_2 \,|| \, \ldots \,|| \, I_{k}$, $k\geq3$;
    \item[($ii$)] if $w=2^n-1$,
$\pi^*$ satisfies the 2iS property, i.e.,
$\pi^*= \Bigl( I_1 \backslash (2n+1) \Bigr) \ ||\  I_2 \  ||\ (2n+1)$,
where $I_1= (n+1, n+2, \ldots, 2n, 2n+1)$ and $I_2= (1, 2, \ldots, n)$.
    \end{enumerate}
\end{theorem}

\vspace*{0.1in}
\section{Properties of the Flow-graph $F[\pi^*]$}
\label{sec:Properties-RPG}

Collberg et al. \cite{CT99,CKCT03} describe several techniques
for encoding watermark integers in graph structures.
Based on the fact that there is a one-to-one correspondence,
say, $\mathcal{C}$, between self-inverting permutations and isomorphism classes
of RPGs, Collberg et al. \cite{CKCT03} proposed a polynomial-time
algorithm for encoding the integer $w$ as the RPG corresponding to
the $w$th self-inverting permutation $\pi$ in $\mathcal{C}$.
This encoding exploits only the inversion property of
a self-inverting permutation and does not incorporate
any other property.

In our codec system {\tt(encode,\,decode)}$_{\tt F[\pi^*]}$
an integer $w$ is encoded as a self-inverting permutation $\pi^*$
using a construction technique which captures into $\pi^*$
important properties such as the odd-one, the bitonic, the block and the range properties.
In this section, we describe the main properties of our reducible permutation
graph~$F[\pi^*]$ produced by the algorithm
{\tt Encode\_SiP.to.RPG}; we mainly focus on the properties of $F[\pi^*]$
derived from permutation~$\pi^*$ and discuss them with respect to resilience to attacks.

\vskip 0.2in 
\subsection{Codec Properties}

In a graph-based watermarking environment, the watermark graph~$G[w]$ should not differ from the graph data structures of real programs.
Important properties are the maximum outdegree of $G$ which
should not exceed two or three, and the existence of
a unique root node so that all other nodes can be reached from it.
Moreover, $G[w]$ should be resilient to attacks against edge and/or
node modifications. Finally, $G[w]$ should be efficiently constructed.

Our watermark graph $F[\pi^*]$ incorporates all the above properties; indeed, the graph $F[\pi^*]$
and the corresponding codec and system $({\tt encode}, {\tt decode})_{\tt F[\pi^*]}$ have
the following properties:

\begin{enumerate}
\item[$\bullet$\,]
{\bf Appropriate graph types}: The graph~$F[\pi^*]$ is a directed graph
on $n^*+2$ nodes with outdegree at most two; that is, it has
low max-outdegree, and thus it matches real program graphs.

\vspace*{0.01in}
\item[$\bullet$\,]
{\bf High resilience}: Since exactly one node of the graph~$F[\pi^*]$ has outdegree~$0$, exactly one has
outdegree~$1$, and the rest have outdegree~$2$,
we can with high probability identify and correct edge
modifications, i.e., edge-flips, edge-additions, or
edge-deletions.
Thus, the graph~$F[\pi^*]$ enables us to correct edge changes (see Section~\ref{subsec:Edge-Modification}).

\vspace*{0.01in}
\item[$\bullet$\,]
{\bf Small size}: The size $|P_w| - |P|$ of the embedded watermark~$w$
is relatively small since the size of the corresponding watermark
graph~$F[\pi^*]$ is $O(n^*)$; in fact, $F[\pi^*]$'s size is
$O(\log_2 w)$ because $n^*=2n+1$ and $n=\lceil \log_2 w \rceil$.

\vspace*{0.01in}
\item[$\bullet$\,]
{\bf Efficient codecs}:
The codec $({\tt encode}, {\tt decode})_{F[\pi^*]}$ has low time and
space complexity; indeed, we have showed that
both the encoding algorithm {\tt Encode\_SiP.to.RPG} and
the decoding algorithm {\tt Decode\_RPG.to.SiP}
require $O(n^*)$ time and space, where $n^*$ is the size of
the input permutation~$\pi^*$ (see Theorems~\ref{thm:theo5.1} and
\ref{thm:theo5.2}).
\end{enumerate}

\noindent
It is worth noting that our encoding and decoding algorithms use
basic data structures and operations, and thus they are easily
implementable.

\begin{figure}[t]
    \hrule\medskip
    \centering
    \includegraphics[scale=0.6]{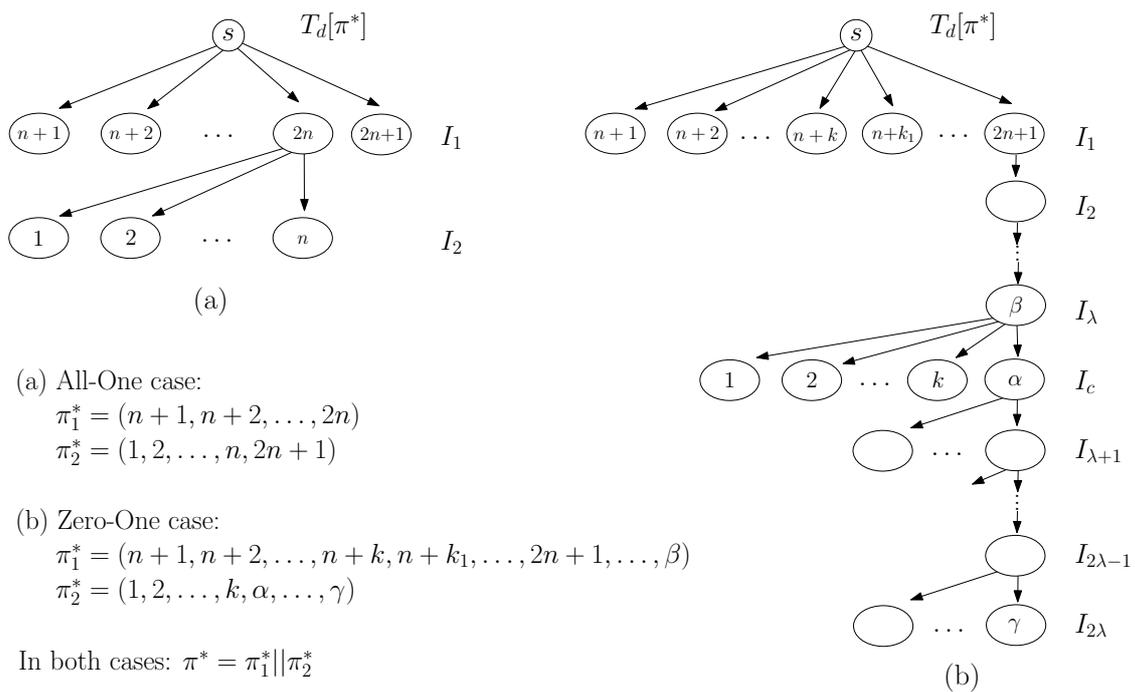}
    \centering
    \smallskip\hrule\medskip
    \caption{\small{(a) All-One case: $\pi^*=(I_1, \ I_2)$,
where $I_1 = \pi_1^* || (2n+1)$ and $I_2 = \pi_2^* \backslash (2n+1)$.
(b) Zero-and-One case: $\pi^* \ = \ (I_1, \ I_2, \ \ldots,
\ I_{\lambda}, \ I_c, \ I_{\lambda+1}, \ I_{\lambda+2}, \ \ldots, \ I_{2\lambda})$.}}
\medskip\medskip
\label{fig:fig5}
\end{figure}

\vskip 0.2in 
\subsection{Structural Properties}

The algorithm {\tt Encode\_SiP.to.RPG} presented
in Subsection~\ref{subsec:SiP-to-RPG}, which encodes
a SiP~$\pi^*$ as a reducible permutation graph~$F[\pi^*]$,
constructs the graph~$F[\pi^*]$ by computing the function
$P(i) = dmax(i)$ for each element $i \in \pi^*$, $1 \leq i \leq n^*$; recall that $dmax(i)$ is the maximum element in the set of all the elements of $\pi^*$ that d-dominate the element~$i$
(see Section~\ref{subsec:Preliminaries}).

Let $I(\pi^*)=[I_1, \ I_2, \ \ldots, \ I_\lambda, \ I_c,
\ I_{\lambda+1}, \ \ldots, \ I_{2\lambda}]$
be the $I$-representation of a SiP $\pi^*$ having the CiS property,
i.e., $\pi^*$ encodes a number $w \in \widehat{R}_n=[2^{n-1}, 2^n-2]$
(case Zero-and-One), and let $i$ be an element of the $k$-th
increasing subsequence~$I_k$ of $\pi^*$, $1 \leq k \leq 2\lambda+1$.
Based on the structure of $I(\pi^*)$, it is easy to see that
$dmax(i)$ is the maximum element of the $(k-1)$-st
increasing subsequence~$I_{k-1}$ of $\pi^*$ which, in turn, is
the last element of $I_{k-1}$; by convention, $I_0=(s)$.
Indeed, in our Example~4.3 where $\pi^*$ encodes the number $w=105$,
we have
$$I(\pi^*) = [(8,9,11,14,15), \ (13), \ (12), \ (1,2,10),
\ (3,7), \ (6), \ (4,5)]$$
and we can easily see that $dmax(5) = dmax(4) = 6$,
$dmax(6)=7$, $dmax(7) = dmax(3) = 10$, $dmax(10)=12$, and so on,
whereas $dmax(15) = dmax(14) = dmax(11) = dmax(9) = dmax(8) = s$.

In the case where $\pi^*$ does not satisfy the CiS property,
i.e., $\pi^*$ encodes the last number in
the range $R_n=[2^{n-1}, 2^n-1]$ (case All-One), we have
a similar property.  In this case, $I(\pi^*)=[I_1, \ I_2]$ where
$I_1 = \pi_1^* || (2n+1)$ and $I_2 = \pi_2^* \backslash (2n+1)$
(see Section \ref{subsec:CiS-Property}).
We observe that for each element $i \in I_1$, $dmax(i)$ is
the maximum element of the increasing subsequence~$I_0=\{s\}$,
whereas for each element $i \in I_2$, $dmax(i)$ is the
\emph{2nd largest} element of the increasing subsequence~$I_1$
of $\pi^*$; see such a SiP~$\pi^*$ along with
its $I$-representation~$I(\pi^*)$ in Example~4.2.

Then, the Dmax-tree $T_d[\pi^*]$ presented in
Subsection~\ref{subsec:SiP-to-RPG}, which is constructed
by exploiting the $dmax$ values, has the following structure:

\begin{enumerate}
\item[$(i)$\,]
{\it All-One case}: The tree~$T_d[\pi^*]$ consists of
the root-node~$s$ at level~$0$ and $2$ more levels:
the $1$st level contains the elements of
the $1$st increasing subsequence $I_1 = \pi_1^* || (2n+1)$;
the $2$nd level contains the elements of the $2$nd increasing subsequence
$I_2 = \pi_2^* \backslash (2n+1)$.
All the nodes of the $2$nd level of $T_d[\pi^*]$ have the node
labeled with the number $2n$ (i.e., the 2nd largest element of $I_1$)
as their parent node.

\vspace*{0.03in}
\item[$(ii)$\,]
{\it Zero-and-One case}: In this general case, the tree~$T_d[\pi^*]$ consists of the root-node~$s$ at level~$0$ and $2\lambda+1$ more levels, where $2\lambda+1$ is the length of the $I$-representation
$I(\pi^*)=[I_1, \,I_2, \,\ldots, \,I_\lambda, \,I_c, \,I_{\lambda+1},
\,\ldots, \,I_{2\lambda}]$ of the SiP $\pi^*$:
the $k$th level contains the elements of
the $k$th increasing subsequence $I_i$ of $I(\pi^*)$.
All the nodes of the $k$th level of $T_d[\pi^*]$ have
as their parent node the node labeled with the maximum element
of the $(k-1)$-st increasing subsequence $I_{k-1}$ of $\pi^*$;
by convention, we consider $I_{0}=\{s\}$.
\end{enumerate}

\noindent
The two different structures of the Dmax-tree~$T_d[\pi^*]$ corresponding to the All-One and the Zero-and-One cases are presented in Figure~\ref{fig:fig5}.

It is worth noting that the very strict structure of
a self-inverting permutation~$\pi^*$ produced by
Algorithm {\tt Encode\_W.to.SiP} enables us to obtain
the increasing subsequences of $\pi^*$ and construct
the watermark graph~$F[\pi^*]$ in time linear in its size and,
more importantly, makes the graph~$F[\pi^*]$ robust and resilient
to attacks.

\vskip 0.2in 
\subsection{Unique Hamiltonian Path}

It is well-known that any acyclic digraph~$G$ has at most one
Hamiltonian path (HP) \cite{CKCT03}; $G$ has one HP if there exists an ordering
$(v_1, v_2, \ldots, v_n)$ of its $n$ nodes such that
in the subgraphs $G_0, G_1, \ldots, G_{n-1}$
the nodes $v_1, v_2, \ldots, v_n$, respectively, are
the only nodes with indegree zero, where $G_0=G$ and
$G_i = G \backslash \{v_1, v_2, \ldots, v_i\}$, $1 \leq i \leq n-1$.
Furthermore, it has been shown that any reducible flow-graph
has at most one Hamiltonian path \cite{CKCT03}.

It is not difficult to see that the reducible permutation
graphs~$F[\pi^*]$ constructed by Algorithm {\tt Encode\_SiP.to.RPG}
have a unique Hamiltonian path HP$(F[\pi^*])$;
it is precisely the path $(u_{n^*+1}, u_{n^*}, \cdots, u_1, u_0)$.
Such a path can be found in time linear in the size of $F[\pi^*]$ by the following algorithm.


\vspace*{0.15in} \noindent Algorithm {\tt Unique\_HP}
\vspace*{-0.04in}
\begin{enumerate}
\item[1.\,]   
Find the unique node, say, $u$, of the graph~$F[\pi^*]$ with outdegree~$1$;
\vspace*{0.01in}
\item[2.\,]   
Perform DFS-search on graph $F[\pi^*]$ starting at node~$u$
and compute the DFS discovery time~$d[v]$ of each node~$v$
of $F[\pi^*]$;
\vspace*{0.01in}
\item[3.\,]   
Return HP$(F[\pi^*])=(u'_0, u'_1, \ldots, u'_{n^*+1})$ where
$(u'_0, u'_1, \ldots, u'_{n^*+1})$ is the ordering of the nodes
of the graph~$F[\pi^*]$ by increasing DFS discovery time~$d[]$,
i.e., $d[u'_i] < d[u'_j]$ for $i<j$, $0 \leq i, j \leq n^*+1$;
\end{enumerate}

\vspace*{0.02in}
\noindent
Since the graph~$F[\pi^*]$ contains $n^*+2$ nodes and $2n^*+1$ edges,
both finding the node of $F[\pi^*]$ with outdegree~$1$ and
performing DFS-search on $F[\pi^*]$ take $O(n^*)$ time and
require $O(n^*)$ space.
Moreover, the ordering of the nodes by their DFS discovery time
is readily obtained during the DFS-traversal.
Thus, we have the following result.

\vspace*{0.04in}
\begin{theorem}\label{thm:theo7.1}
Let $F[\pi^*]$ be a reducible permutation graph of size $O(n^*)$
constructed by algorithm {\tt Encode\_SiP.to.RPG}.
The algorithm {\tt Unique\_HP} correctly computes the unique
Hamiltonian path of $F[\pi^*]$ in $O(n^*)$ time and space.
\end{theorem}

\vspace*{0.1in}
\section{Detecting Attacks}
\label{sec:Detecting-Attacks}

\noindent
In this section, we show that the malicious intentions of
an attacker to prevent our system from returning
the correct watermark value by modifying node-labels or edges
of the graph $F[\pi^*]$ can be efficiently detected in most cases.

\vskip 0.2in
\subsection{Node-label Modification}
\label{subsec:Node-label-Modification}

\noindent
By construction, our reducible permutation graph~$F[\pi^*]$
is a node-labeled graph on $n^*+2$ nodes, where $n^*$ is
the length of $\pi^*$. The labels of $F[\pi^*]$
are numbers of the set $\{0, 1, \ldots, n^*+1\}$, where
the label $n^*+1$ is assigned to header node $s=u_{n^*+1}$,
the label~$0$ is assigned to footer node $t=u_{0}$, and
the label $n^*+1-i$ is assigned to the $i$th node~$u_{n^*+1-i}$ of the body of $F[\pi^*]$,
$1 \leq i \leq n$.

A label modification attacker may perform swapping of the labels
of two nodes of $F[\pi^*]$, altering thus the value of the label of a node,
or even removing all the labels of the graph $F[\pi^*]$ resulting
in a node-unlabeled graph. Since the decoding of the watermark~$w$
relies on the labels of the flow-graph $F[\pi^*]$
(see algorithm {\tt Decode\_RPG.to.SiP}),
it follows that our codec system
$({\tt encode}, {\tt decode})_{\tt F[\pi^*]}$ is susceptible
to node-label modification attacks.

Therefore, it is important to have a way to decode
the watermark~$w$ efficiently from $F[\pi^*]$ without relying on
its labels. The correct labels can be easily obtained in $O(n^*)$ time
and space thanks to the unique Hamiltonian path HP$(F[\pi^*])$
and the fact that the node-labels of $F[\pi^*])$ are encountered along HP$(F[\pi^*])$
in decreasing order.
Thus, we have the following result.

\vspace*{0.04in}
\begin{lemma}\label{lemma:lemm8.1}
Let $F[\pi^*]$ be a reducible permutation graph of size $O(n^*)$
produced by algorithm {\tt Encode\_SiP.to.RPG}
and let $F'[\pi^*]$ be the graph resulting from $F[\pi^*]$
after having modified or deleted the node-labels of $F[\pi^*]$.
The flow-graph~$F[\pi^*]$ can be reconstructed from $F'[\pi^*]$
in $O(n^*)$ time and space.
\end{lemma}

Moreover, since the number of nodes of the graph~$F[\pi^*]$
is odd, any single node modification in $F[\pi^*]$,
i.e., a node-addition or a node-deletion, can be easily identified.

\begin{table}[t]
\begin{center}
\scalebox{0.8}{
\begin{tabular}{|c||*{6}{c|}}
 \hline
 \multicolumn{7}{|c|}{\bf True-incorrectness Results} \\
 \hline
\diagbox[width=7em]{Range}{Edges}
&$1$&$2$&$3$&$4$&$5$&$6$
\\\hline\hline
$R_4$ & \ \ \ \ \ 0 \ \ \ \ \ & \ \ \ \ \ 0 \ \ \ \ \ & \ \ \ \ \ 0.0005 \ \ \ \ \ & \ \ \ \ \ \ \ 0.0002 \ \ \ \ \ \ \ & \ \ \ \ \ \ \ 0.00009 \ \ \ \ \ \ \ & \ \ \ \ \ \ \ 0.00003 \ \ \ \ \ \ \  \\\hline
$R_5$       &$0$    & $0$ & $0.0003$  &   $0.00005$                 & $0.7 \times 10^{-5}$      & $0.2 \times 10^{-5}$  \\\hline
$R_6$       &$0$    & $0$ & $0.0002$  &   $0.00001$                 & $0.1 \times 10^{-5}$      & $0.2 \times 10^{-6}$  \\\hline
$R_7$       &$0$    & $0$ & $0.0001$  &   $0.7 \times 10^{-5}$    & $0.9 \times 10^{-6}$      & $0.1 \times 10^{-6}$  \\\hline
$R_8$       &$0$    & $0$ & $0.00008$ &   $0.4 \times 10^{-5}$    & $0.5 \times 10^{-6}$      & $0.1 \times 10^{-7}$  \\\hline
$R_9$       &$0$    & $0$ & $0.00005$ &   $0.2 \times 10^{-5}$    & $0.1 \times 10^{-6}$      & $0.1 \times 10^{-7}$  \\\hline
$R_{10}$    &$0$    & $0$ & $0.00003$ &   $0.9 \times 10^{-6}$    & $0.8 \times 10^{-7}$      & $0.1 \times 10^{-8}$  \\\hline
\end{tabular}
}
\end{center}
\caption{\small{The ratio of experiments in which the flow-graph~$F[\pi^*]$ encoding
the number $w \in R_n=[2^{n-1}, 2^n-1]$ was found to be true-incorrect
after $k$ edge-modifications, where $1 \leq k \leq 6$ and
$n=4, 5, \ldots, 10$.}}
 \label{tab:table2}
\end{table}

\vspace*{0.1in}
\subsection{Edge Modification}
\label{subsec:Edge-Modification}

\noindent
We next argue that we can decide, in nearly all cases,
whether the reducible permutation graph~$F[\pi^*]$ produced
by our codec system has suffered an attack on its edges.

Let $F[\pi^*]$ be a flow-graph which encodes the integer $w$ and
let $F'[\pi^*]$ be the graph resulting from $F[\pi^*]$ after
an edge modification. Then, we say that $F'[\pi^*]$ is either a {\it false-incorrect} or a {\it true-incorrect} graph:

\begin{enumerate}
\item[$\bullet$\,] $F'[\pi^*]$ is {\it false-incorrect}
if our codec system fails to return
an integer from the graph $F'[\pi^*]$, whereas
\item[$\bullet$\,] $F'[\pi^*]$ is {\it true-incorrect}
if our system extracts from $F'[\pi^*]$ and returns
an integer $w' \neq w$.
\end{enumerate}

\noindent
Since the SiP properties of the permutation $\pi^*$ which compose
the 4-Chain property
(see Subsection~\ref{subsec:4chain}) are incorporated in
the structure of the reducible permutation graph~$F[\pi^*]$,
it follows that the graph~$F'[\pi^*]$ resulting from $F[\pi^*]$
after any edge modification is false-incorrect if
at least one of the SiP properties does not hold.

We experimentally evaluated the resilience of
the watermark graph~$F[\pi^*]$ in edge modifications.
To this end, we have produced graphs~$F[\pi^*]$
on $2n+3 = 11, 13, \ldots, 23$ nodes (i.e., $n = 4, 5, \ldots, 10$)
and computed the ratio of the experiments in which
the graph~$F_k[\pi^*]$ was found to be true-incorrect, where
$F'_k[\pi^*]$ is the graph resulting from $F[\pi^*]$ after
the modification of $k$ edges, $1 \leq k \leq 6$.
More precisely, our experimental study is based on the following algorithmic scheme:

\medskip
for each $n = 4, 5, \ldots, 10$:
\vspace*{-0.05in}
\begin{itemize}
\item[]
for each number~$w_i$ in the range $R_n = [2^{n-1}, 2^{n}-1]$:
\vspace*{-0.05in}
\begin{itemize}
\item[1.\,]
encode the number $w_i$ as a SiP~$\pi^*$ of
length~$n^*=2n+1$ using Algorithm {\tt Encode\_W.to.SiP}
and then encode $\pi^*$ as a reducible permutation graph~$F[\pi^*]$
of order~$n^*+2$ using Algorithm {\tt Encode\_SiP.to.RPG};

\item[2.\,]
for each $k = 1, 2, \ldots, 6$:

\begin{itemize}
\item[]
repeat the following two steps
$N_n = \lceil \frac{(2n+1)}{9} \times 10^5 \rceil$ times
in order to randomly modify $k$ edges of the graph~$F[\pi^*]$
and count the number~$N(w_i, n, k)$ of times that
the modified graph~$F'_k[\pi^*]$ is true-incorrect;

\begin{itemize}
\item[a.\,] randomly select $k$ edges~$(u_x,u_y)$ and
$k$ nodes~$u_z$ from the graph~$F[\pi^*]$;
\item[b.\,] delete the $k$ edges~$(u_x,u_y)$ and add the $k$
edges~$(u_x,u_z)$ thus yielding the mo\-di\-fied graph~$F'_k[\pi^*]$;
\end{itemize}  
\end{itemize}  
\end{itemize}  
\end{itemize}  
%

\vspace*{-0.05in}

for each $n = 4, 5, \ldots, 10$ and each $k = 1, 2, \ldots, 6$:
\vspace*{-0.05in}
\begin{itemize}
\item[] print the ratio
$\left( \sum_{i = 2^{n-1}}^{2^{n}-1} N(w_i, n, k) \right) /
\left( 2^{n-1} \times N_n \right)$ of true-incorrect cases.
\end{itemize}  

\noindent Note that as $n$ increases, the size of the reducible permutation
graph~$F[\pi^*]$ increases linearly;
thus, for each $w_i \in R_n = [2^{n-1}, 2^{n}-1]$, we repeat
the edge modification experiment $N_n$ times, where $N_n$
depends linearly on $n$.

\begin{figure}[t]
    \hrule\medskip\smallskip
    \centering
    \includegraphics[scale=0.57]{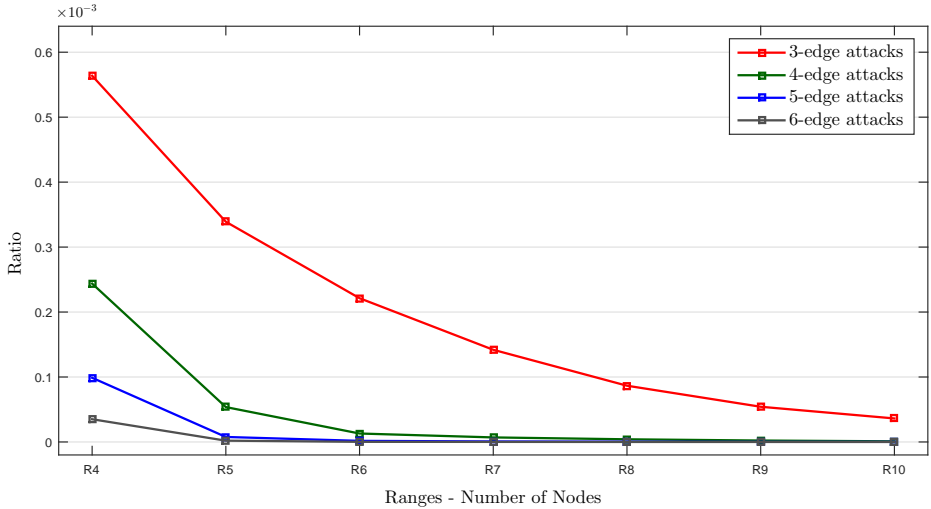}
    \centering
    \medskip\smallskip\hrule\medskip
    \caption{\small{The ratio of experiments in which the RPG $F[\pi^*]$ was found to be true-incorrect after a modification of $3$, $4$, $5$, and $6$ edges.}}
\smallskip\smallskip
\label{fig:Plot-1}
\end{figure}

The experimental results show that the computed ratio of
true-incorrect cases is really small and falls dramatically
as $n$ gets larger (see Table~\ref{tab:table2}
and Figure~\ref{fig:Plot-1}); in fact, no true-incorrect case
is possible if at most 2 edges are modified \cite{BBMPS13}.
Thus, we can decide with high probability whether
our reducible permutation graph~$F[\pi^*]$ has suffered an attack
on its edges.

\vspace*{0.1in}
\section{Concluding Remarks}
\label{sec:Concluding-Remarks}

In this paper we proposed an efficient and easily implementable
codec system for encoding watermark numbers as graph structures.
Our codec $({\tt encode}, {\tt decode})_{F[\pi^*]}$ system
incorporates several important properties and characteristics
which make it appropriate for use
in a real software watermarking environment.
The reducible permutation flow-graph~$F[\pi^*]$ resembles
the graph data structures of real programs since its maximum outdegree
does not exceed two and it has a unique root node.
Additionally, the self-inverting permutation~$\pi^*$ and
the flow-graph~$F[\pi^*]$ encompass important structural properties,
which make our codec system resilient to attacks;
indeed, the graph~$F[\pi^*]$ is highly insensitive to
small edge-changes and fairly insensitive to small node-changes
of $F[\pi^*]$.
Finally, we point out that our codec system
has very low time and space complexity which is $O(n)$,
where $n$ is the number of bits in the binary representation of
the watermark integer~$w$.

In light of the two main data components of our codec system,
i.e., the permutation~$\pi^*$ and the graph~$F[\pi^*]$,
it would be very interesting to come up with new efficient
codec algorithms and structures exhibiting an improved behavior
with respect to resilience to attacks; we leave it as
an open question.
Another interesting question with practical value is
whether the class of reducible permutation graphs
can be extended so that it includes other classes of graphs
with structural properties capable to efficiently encode
watermark numbers.

Finally, the evaluation of our codec algorithms and structures
under other watermarking measurements in order to obtain
detailed information about their practical behavior is
an interesting problem for future study.

\end{document}